\documentclass{article}

\usepackage{arxiv}
\usepackage{macros/macros}
\usepackage{microtype}

\begin{document}
\pagestyle{fancy}
\thispagestyle{empty}
\fancyhf{}

\newcommand{\runningtitle}{AI-Generated Non-Consensual Intimate Imagery Is Configured, Not Built}
\newcommand{\runningauthors}{Kieserman et al.}

\fancyhead[L]{\sffamily\footnotesize\ifodd\value{page}\runningtitle\fi}
\fancyhead[R]{\sffamily\footnotesize\ifodd\value{page}\else\runningauthors\fi}
\renewcommand{\headrulewidth}{0pt}

\title{AI-Generated Non-Consensual Intimate Imagery is Configured, Not Built: How General-Purpose AI Technology Enables Sexual Violence}

\author{\begin{tabularx}{\textwidth}{@{} *{3}{>{\centering\arraybackslash}X} @{}}
{ Julia B. Kieserman}
&
{ Ashley Phoebe Ishoel}
&
{ Damon McCoy}
\\
{\small \normalfont New York University\footnotemark[4]}
&
{\small \normalfont Indiana University}
&
{\small \normalfont New York University}
\\
{\footnotesize\texttt{julia.kieserman@nyu.edu}}
&
{\footnotesize\texttt{ashleyishoel@proton.me}}
&
{\footnotesize\texttt{mccoy@nyu.edu}}
\\[1em]
{ Elissa M. Redmiles}
&
{ Lucy Qin}
&
{ Jaron Mink}
\\
{\small \normalfont Georgetown University}
&
{\small \normalfont Georgetown University}
&
{\small \normalfont Arizona State University}
\\
{\footnotesize\texttt{elissa.redmiles@georgetown.edu}}
&
{\footnotesize\texttt{lucy.qin@georgetown.edu}}
&
{\footnotesize\texttt{jaron.mink@asu.edu}}
\end{tabularx}}

\maketitle
\footnotetext[4]{This work was conducted in part while the author was a visiting scholar at Georgetown University.}

\begin{abstract}
AI-generated non-consensual intimate imagery (AIG-NCII) is an increasingly prevalent form of online abuse, distinguished from AI-generated sexual content (AIG-SC) by the \emph{sexualized} depiction of a real person's \emph{identity} without their consent. To understand how offenders may benefit from the tools and resources shared across the broader AIG-SC community, we analyzed \numGuidestotal R+ guides on \civitai, a popular platform known to host a disproportionate amount of AIG-SC.

We find that AIG-SC guides discuss how to create content generation workflows composed of sexual and identity concepts, the building blocks of AIG-NCII, including one guide that explicitly outlines an AIG-NCII workflow. We identify that ultimately AIG-NCII is enabled by configuration rather than a distinct set of technical tools. We map AIG-NCII
threat models across basic and advanced AIG-SC workflows to demonstrate how multiple dual-use
technical components can compose NCII outputs in an environment with little to no discussion of ethical implications or responsible use. We conclude by noting that, as most workflows rely on open-weight models, existing technical defenses against AIG-NCII are limited. However, our findings offer a starting point for evaluation of AI-based models and products against realistic attacker behavior required by new EU regulation and suggest that efforts to develop safeguards at the model- and product-levels should focus on generation of \emph{identity} content.

\end{abstract}

\noindent\fbox{\parbox{\dimexpr\linewidth-2\fboxsep-2\fboxrule\relax}{\small \textbf{Content Warning:} This paper uses medical anatomical terminology to discuss human image generation.
    This paper also discusses sexual content and identity-based harassment, including the non-consensual creation of AI-generated intimate imagery. This includes first-hand accounts from offenders who have created NCII.
  }}

\section{Introduction}
AI-generated non-consensual intimate imagery (AIG-NCII) is an increasingly common~\cite{hawkinsDeepfakesDemandRise2025} form of image-based sexual abuse (IBSA)~\cite{mcglynnRevengePornContinuum2017} in which offenders generate an image that depicts a real, identifiable person in a sexualized manner.
Victim-survivors of IBSA may report severe emotional distress that can lead to depression, anxiety, or post-traumatic stress disorder~\cite{batesRevengePornMental2017, huberShadowMeOld2023, henryDevastatingItBroke2023, martin2022image}. These harms are exacerbated by content removal processes that can be lengthy and difficult to navigate~\cite{huberNonconsensualIntimateImage2024, qiweiPlatformsCrimeScene2026, beccabranumRapidResponseBuilding2025}.

Although the non-consensual creation of intimate content predates AI, earlier tools, such as photo editing software (e.g., Photoshop)~\cite{chaMyBodyNot2026}, did not scale at the same pace as the AI-generated methods available today. In particular, journalists, academics, and lawmakers have  focused their attention on applications that explicitly advertise NCII (e.g., ``nudification'' apps) since these apps are easy to use, cheap (or free), and relatively easy to discover~\cite{gibson2025analyzing}.
Accordingly, U.S. regulators have passed laws banning nudification applications~\cite{minnesotaNudificationLaw} and mandating that social media platforms remove NCII (whether AI-generated or not) within 48 hours of a victim-survivor's content removal request~\cite{tedcruzS146TAKEIT2025, code6851}. Applications like AI-powered face-swapping apps or image-editors are increasingly identified as ``dual-use,'' meaning they support both abusive and non-abusive workflows~\cite{daffalla2026dual, minkUnlimitedRealmExploration2026, williams2025there, emanuelmaibergInstagramsNudifyAds2024}. As more creation methods are identified, the E.U. now requires AI model and application developers to proactively test for ``reasonably foreseeable misuse''~\cite{euAIAct} by AIG-NCII offenders.

AI applications are typically powered by a workflow: a series of technical components linked together to make effective use of a ``base'' generative AI model. Although researchers have identified instances where dual-use technical components used in these workflows may facilitate abuse~\cite{thielGenerativeMLCSAM2025, hawkinsDeepfakesDemandRise2025, han2025characterizing, cui2026celebrities},
there has been no systematic evaluation of how the larger workflows created from these components facilitate abuse. Just as full-system evaluations (i.e. workflows) improve coverage when testing for privacy side-channels in machine learning systems~\cite{debenedetti2024privacy} and software fuzzing~\cite{bernhard2024darthshader,ruge2020frankenstein}, models or applications must be evaluated for AIG-NCII abusability in the workflows in which they will reasonably be used~\cite{redmiles2024nist}, not evaluating a base model ``in a vacuum''~\cite{debenedetti2023privacy}.

Subsequently, we present a first analysis of the AIG-NCII abusability of AI visual-media generation workflows. We do so through a qualitative analysis of public guides for generating AI visual media shared on \civitai, a popular platform known to host a disproportionate amount of sexual content~\cite{wagnerPerpetuatingMisogynyGenerative2025, weiExploringUseAbusive2024}\footnote{Because \civitai{} disproportionately hosts models used to create sexual content, our analysis of its guides focuses primarily on the potential for creating AIG-NCII that is sexual. As prior work has noted~\cite{minkUnlimitedRealmExploration2026}, AI-generated sexual content may be more difficult to generate than other forms of content. Therefore, there may be even lower barriers to creating other forms of AIG-NCII. We rely on the Take It Down Act's definition of ``intimate visual depiction'' to anchor our analysis~\cite{tedcruzS146TAKEIT2025}. However, we recognize that legislative definitions can be imperfect. In general, definitions for NCII are highly contextual and personal. For example, NCII can also include clothed images that imply a sexual or romantic relationship, depending on cultural context~\cite{chaynExplicitHarmsNonExplicit2026, batoolExpandingConceptsNonConsensual2024}.} and to be a hub for resources shared by the AI-generated sexual content (AIG-SC) community, who generate content for creative and sexual expression of specific kinks, aesthetics, or fictional personas~\cite{minkUnlimitedRealmExploration2026, doringExperiencesAIGeneratedPornography2025}.
Specifically, our analysis seeks to answer the following research questions:
\begin{enumerate}[label=\textbf{RQ\arabic*}, labelindent=0pt, leftmargin=*, align=left, nosep]    \item How are AIG-NCII relevant goals discussed in AIG-SC guides? \label{rq:aig-ncii}
    \item What is the architecture of AIG-SC workflows --- and how are the NCII-enabling concepts of (a) sexual (\sexualrisklabel) and (b) identity (\identityrisklabel) risks enabled?
    \label{rq:architecture-and-risks}
    \item How do guides discuss the governance and ethics of introducing NCII-enabling concepts? \label{rq:aig-sc-guides-ethics}
\end{enumerate}

We find that increasingly UI-bound~\cite{freedStalkersParadiseHow2018} workflows shared on \civitai{} can assist AIG-NCII offenders in
creating more sophisticated and custom depictions of \emph{real sexualized identities} with unregulated components that render AIG-NCII creation seemingly invisible.

We identify that AIG-SC guides discuss generating specific, consistent personas in the context of sexual content; in some cases (\intimatexfictionalidentityGuidesCount) this includes generating sexualized depictions of fictionalized identities or, in one instance, of a non-fictional identity (Section \ref{sec:ncii-goals}). We outline two workflows commonly used by AIG-SC creators (\emph{basic} and \emph{advanced}) and enumerate the ways in which guides teach readers to configure identity and sexual content such that offenders can produce AIG-NCII directly (Section \ref{subsec:architecture}).
Through this analysis, we both validate that previously identified components for NCII-generation, such as the training and use of custom LoRAs~\cite{ortegaAILovesBoobies2026, medeirosCharacterizingResourceSharing2026, hawkinsDeepfakesDemandRise2025}, remain a commonly used tool in the AIG-SC community (\iref{irisk:loras-of-people}, \sref{srisk:lora-sexual-modifications}) and identify new tactics, like ``likeness prompting'' (\iref{irisk:prompting-persona}) and LLM (Large Language Model)-assisted upskilling (\sref{srisk:llm-upskilling}-\sref{srisk:faceswapping-sexual}, \iref{irisk:llm-upskilling}-\iref{irisk:faceswapping-identity}), taught across AIG-SC guides, that offenders use to support AIG-NCII content generation.
While technical defenses do exist to prevent some of these workflow configurations from generating abusive content, \civitai{} guides discuss how to run workflows locally on commercially available hardware and primarily rely on open-weight models, especially fine-tuned derivatives (\finetunedGuidesCount), meaning that defenses can easily be circumvented.

Through this lens, NCII can be understood as a customization of general-purpose content workflows. This suggests that the line between generating AIG-SC and AIG-NCII is a sociotechnical one. While \civitai{} was previously found to host and disseminate abusive content~\cite{weiExploringUseAbusive2024, ghosh2026marketplace, hawkinsDeepfakesDemandRise2025}, in
response to the introduction of laws~\cite{takeItDown, EUai} and pressure from payment processors~\cite{maiberg25_civitai_cutoff}, the platform
removed models and images of non-consensually generated sexual content that depicts the identity of a real person~\cite{civitaiPolicy, maiberg25_civitai_people_ban}. As this work shows, such measures are insufficient.  \civitai{} is still used to share generalizable, open-weight resources and knowledge across platforms~\cite{qinThoughtYouWere2026} that further construct the broader AIG-NCII creation ecosystem and
\civitai{} community members infrequently discuss the boundary that separates legitimate AIG-SC from AIG-NCII, in some cases explicitly mentioning ways to evade platform restrictions (Section \ref{sec:boundaries}). We conclude by offering possible means of adding friction to curtail AIG-NCII (Section~\ref{sec:discussion}).

\section{Background \& Related Literature}

\header{AI-Generated Sexual Content}
\label{subsec:aig-sc}
There are a growing number of image generation services that produce AI-generated sexual content (AIG-SC) based on text and/or image prompts with minimal technical skills~\cite{lapointePresentFutureAdult2025}. Further, prior work has demonstrated the existence of large online communities that support AIG-SC creation~\cite{doringExperiencesAIGeneratedPornography2025,minkUnlimitedRealmExploration2026}.
Creators in this space have a variety of motivations for creating sexual content, including sexual expression, creative expression, and technical interest in content creation~\cite{minkUnlimitedRealmExploration2026}. \civitai{} is a popular platform used to host, create, and disseminate knowledge on the creation of AI-generated content, including fine-tuned models, images/videos, and user-contributed articles (the focus of this study), in which creators share instructional information on how to create a range of custom content using local or cloud-based tools. \civitai{} is known to host a disproportionate amount of sexual content, which receives more engagement than other forms of content~\cite{wagnerPerpetuatingMisogynyGenerative2025, weiExploringUseAbusive2024}.
However, the technical processes for creating AIG-SC can be misappropriated to create AI-generated non-consensual intimate images (AIG-NCII). Subsequently, the advice and technical resources disseminated by communities of AIG-SC creators can indirectly benefit offenders of AIG-NCII. In the absence of robust legal protections or even community-based governance, there are direct opportunities for this ecosystem to facilitate the creation of AIG-NCII.

\header{AIG-NCII}
As generating AIG-SC has gotten easier, so too has generating AIG-NCII. Apps like nudification tools are easy to use --- ``digitally undress anyone with the click of a button''~\cite{jasminemithaniAdvocates} --- and discoverable with mainstream search engines~\cite{williams2025there, gibson2025analyzing, samanthacoleNewResearchShows2025}.
Even apps that do not explicitly advertise the creation of AIG-NCII can be re-purposed to do so, such as generic face-swapping apps, image-editing apps, or jailbroken image generation apps, which often lack sufficient safeguards to prevent the creation of AIG-NCII~\cite{daffalla2026dual, minkUnlimitedRealmExploration2026}. Similarly, AI chatbot services include features that allow for the reproduction of likeness, in some cases, in tandem with the creation of nude/lewd content~\cite{brighamExaminingRisksAI2026, alexiosmantzarlisAICompanionApps, gracebrighamHowAICompanion2026}.
\civitai, the focus of the study, has consistently been found to host and disseminate abusive content, including both models that output non-consensual intimate media and media that itself infringes copyright~\cite{weiExploringUseAbusive2024, ghosh2026marketplace, hawkinsDeepfakesDemandRise2025}.

To begin to reduce the ease of access to AIG-NCII creation, regulation and platform attention have turned to nudification apps. Minnesota became the first U.S. State to pass a ban on websites and apps that use generative AI to ``nudify'' individuals depicted in user-uploaded images~\cite{jasminemithaniMinnesotaPassesNations2026}. At a platform level, Apple and Google have banned nudification apps in their app stores and are working to remove such apps~\cite{emanuelmaibergAppleRemovesNonconsensual2024, lolamurtiTechAppleGoogle2026}, with varying levels of efficacy~\cite{techtransparencyprojectAppleGoogleAre2026}.
The introduction of the U.S. ``TAKE IT DOWN Act''~\cite{tedcruzS146TAKEIT2025, code6851}, which went into effect in May 2026 and ``criminalizes the non-consensual publication of intimate images'' has led to the deplatforming of sites like Mr.Deepfakes, which hosted explicit-NCII workflows~\cite{han2025characterizing, cuevas2026deepfake}.

In response to emerging regulation, \civitai{} introduced input filtering in 2023 to deter the creation of NCII depicting known identities (i.e. celebrities). However, this
did not address the existence of NCII more broadly, particularly for lesser-known identities~\cite{emanuelmaibergAIPornMarketplace2023}; \civitai{} continued to host customized models that could be used to generate content depicting celebrities, influencers, athletes, adult content performers, and others \cite{emanuelmaibergAIPornMarketplace2023}. In 2025,
facing pressure from payment processors~\cite{maiberg25_civitai_cutoff}, \civitai{} implemented an additional platform policy that bans models that depict a real person~\cite{civitaiPolicy, maiberg25_civitai_people_ban}, relying on self-attestation for ``unknown'' identities (Appendix \ref{app:self-attestation}).

These efforts strategically target an important (and accessible) part of the AIG-NCII creation ecosystem. However, they do not address the underlying generative AI techniques that underpin these services, nor the ways in which AIG-SC technical workflows can be configured for abuse
~\cite{han2025characterizing, wiseMajorDeepfakePorn2025, dawoudUndergroundMainstreamMarketplaces2026}. Doing so requires a deeper understanding of the systematic ways in which workflows can be configured to support AIG-NCII. To that end, we present a technical analysis of how the two fundamental building blocks of AIG-NCII, sexual and identity content, are introduced in \civitai{} articles.

\section{Methodology}
\label{sec:methods}
To analyze how AIG-NCII relevant goals are discussed in guides (\ref{rq:aig-ncii}), how the architecture of AIG-SC workflows, as taught by guides enable those goals (\ref{rq:architecture-and-risks}), and how guides discuss boundaries around introducing NCII-enabling content (\ref{rq:aig-sc-guides-ethics}), we collected and analyzed \numGuidestotal AIG-SC guides across \civitai.
We discuss ethical considerations in the Appendix.

\subsection{Terminology}
\label{sec:terminology}
We begin by defining the terminology used in this work.

\header{Personas \& Identities}
``Personas'' are depictions of a human (or human-like), anthropomorphic subject.
Leveraging existing legal definitions of the expression of characters and individuals~\cite{publicity_law, schreyer2015overview, takeItDown}, we define an ``identity'' as the distinctive, distinguishable characteristics that make up an individual persona;
this does not depend on a content's style (e.g., animated or photorealistic).
While the visual expression of real individuals and fictitious personas may be subject to different legal branches and constraints (e.g., publicity~\cite{publicity_law}, copyright \& trademark~\cite{schreyer2015overview}, or criminal~\cite{takeItDown} law in the U.S.),
we define identities as the consistent representation of both real and fictitious personas.
We do this, as the objective (from a technical perspective) of representing an existing identity, real or fictitious, in AI-generated content is extremely similar; as such, we suspect (and in our results confirm) that the technical requirements and methods for both are often the same.

\header{Sexual Content}
To avoid subjective interpretation that overlooks contextual nuance and reinforces societal biases often present in vague definitions used by platforms and computer science research~\cite{riccioExposedErasedAlgorithmic2024a, gehlTrainingComputersSee2017, princessacintaqiaStopNonconsensualUse2025},
we clearly define \textit{sexual content} as content that is either ``intimate'' or ``sexually-relevant.''

In line with U.S. criminal codes pertaining to NCII~\cite{tedcruzS146TAKEIT2025, code6851},
we define content as \textit{intimate} if it depicts uncovered genitals, pubic area, anus, breasts, bodily fluids, sexual acts, or BDSM\footnote{Bondage, Discipline, Dominance, Submission, Sadism, and Masochism} acts~\cite{code6851}.
We define content as \textit{sexually-relevant} when it includes concepts that do not meet the threshold of `intimate,' but often appear alongside or within intimate content, or are commonly seen as provocative through conventional Western gazes of sexuality.
This includes the depiction of
(a) sexually-suggestive poses of personas,
(b) human anatomy or its motion, and
(c) undergarments or lingerie.
Both our definitions (and the law) make exceptions for the use of nudity (e.g., genitals, anus, breasts) for educational purposes; however, no guides clearly held this context.

\begin{figure}[t]
    \centering
    \includegraphics[width=0.5\linewidth]{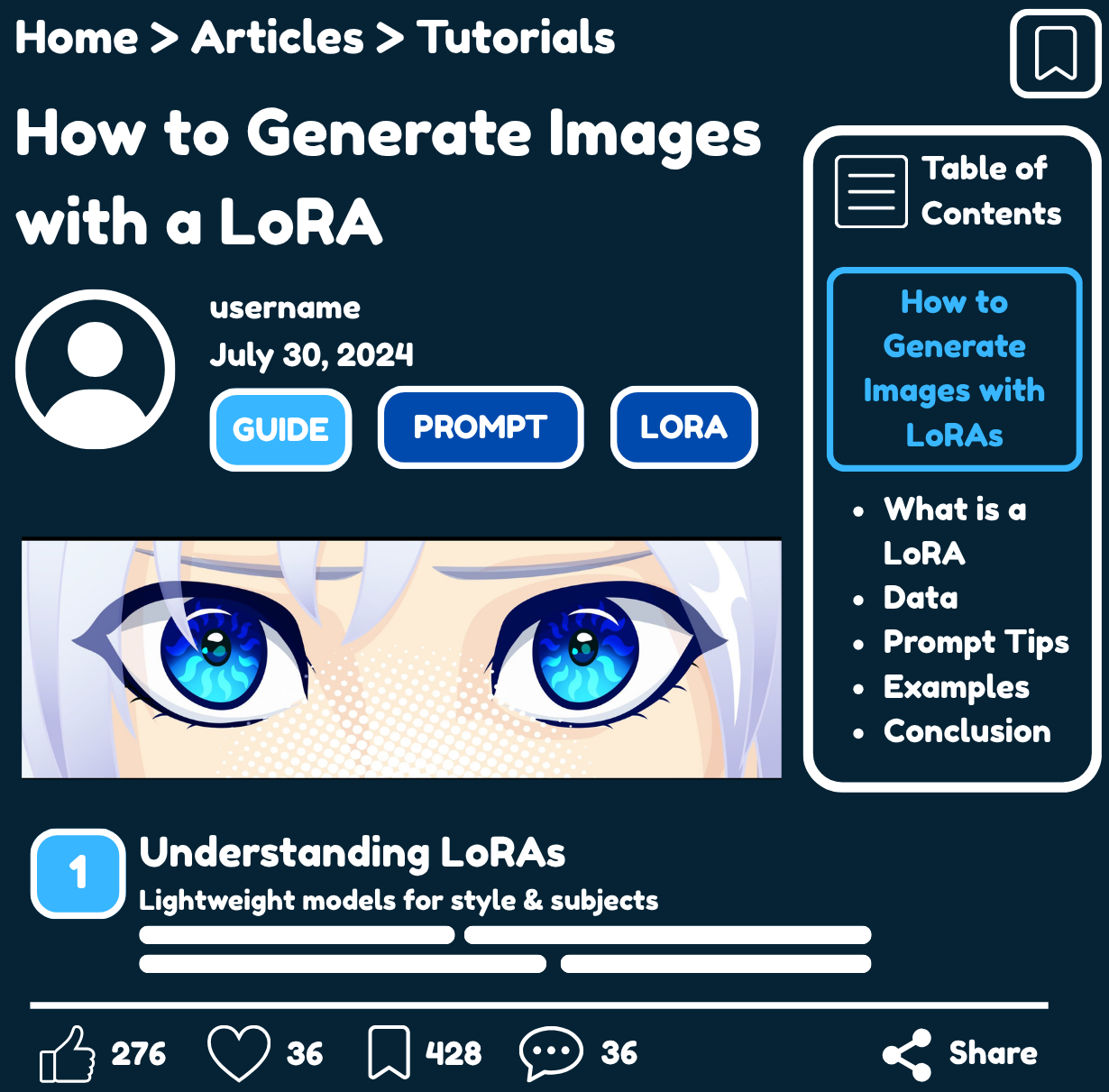}
    \caption{\textmd{\textbf{Stylized Rendering of a \civitai{} Article} --- {\small Images, Text, and Step-by-Step Walkthrough Guide Users in Content Generation Workflows. Stock image credit: Natalia Scharoikina}}}
    \label{civitai-guide-example}
\end{figure}

\subsection{Data Collection}
\label{sec:methods-datacollection}

To obtain an overview of popular, modern methods for generating AIG-SC content, we analyze \numGuidestotal user-generated articles, publicly posted between November 2024--May 2026 on \civitai.
As shown in Figure \ref{civitai-guide-example}, these articles are text-based posts that may also include example outputs, or relevant linked resources (data, models, or additional articles).
Users can bookmark articles and search guides by \# of total bookmarks.

As \civitai{} removed models and images that generate NCII in 2025~\cite{civitaiPolicy, maiberg25_civitai_people_ban}, all guides within \civitai{} should (in theory) not directly enable NCII,
and, following \civitai{} policy, should not include any real identities (sexual or not).
As such, this set of articles is well-curated to assess whether these measures are sufficient to prevent AIG-NCII from being generated by the AIG-SC infrastructure.

\begin{table}[t]
\centering

\rowcolors{2}{gray!15}{white}
\begin{tabular}{lrr}
\toprule
\textbf{Category} & \textbf{Collected} & \textbf{Coded} \\
\midrule
Workflows         & 32 & 31 \\
Generation        & 56 & 49 \\
Tools             & 12 & 9 \\
Training          & 28 & 20 \\
Resource          & 29 & 14 \\
Comparative Study & 15 & 10 \\
Video Generation  & 1 & 1 \\
Data Prep         & 5 & 5 \\
ML Research       & 0 & 0\\
\textbf{Total} & \textbf{178} & \textbf{\numGuidestotal} \\
\bottomrule
\end{tabular}
\caption{\textmd{\textbf{\civitai{} Dataset} --- {\small Guides by Platform Category}}}
\label{tab:civitai_guides}
\end{table}

\header{Guide Criteria}
To ensure a replicable methodology, we limited our collection to articles with the following criteria:

\begin{enumerate}[nosep,leftmargin=*,noitemsep,label=\textit{C\arabic*.}]
   \item \textit{Guides:}
    \civitai{} articles are assigned a content category.
    Articles were considered \textit{guides} if they belonged to one of nine categories we identified as instructive (Table \ref{tab:civitai_guides}).

 \item \textit{Recent:} Published between November 2024--May 2026.

 \item \textit{Adult Rating:} Guides are labeled by \civitai{} and authors for their audience appropriateness;

 we consider adult guides as being rated R, X, or XXX (see Appendix~\ref{app:classification-system}).

 \item \textit{Accessible To Team:} Guides were accessible to the authoring team, if they contained written English text.

  \item \textit{Popular:}
  As a watermark of popularity and perceived utility, we
  considered a guide `popular' if it was ``bookmarked'' (i.e., saved) by at least 20 users.\footnote{We make one exception for ``Generation'' category, which held many more guides compared to all other categories. As such, we collected guides until we reached thematic saturation rather than collecting all guides.}
  The bookmarks for our data had a median of 65, and ranged from 20--1,800.

\end{enumerate}

\header{Guide Collection}
As shown in Table~\ref{tab:civitai_guides}, we collected 178 publicly posted AIG-SC guides across nine article categories and analyzed them until we reached thematic saturation (\numGuidestotal guides).
Given the legal and ethical concerns related to analyzing public (sexual) content, we opted to perform a manual (i.e., non-automated) collection of AIG-SC guides.
This collection was limited to PIs of the work, and vetted for abusive or disturbing content before being passed onto others in the team.
This manual vetting helped identify cases of NCII or child sexual abuse material (CSAM), prevent their storage, and report them to authorities (\civitai{}/NCMEC~\cite{nationalcenterformissing&exploitedchildrenNationalCenterMissing2026}).
We discuss the detailed procedures, reporting, and ethics of this work in \textbf{Ethical Considerations}.

\subsection{Qualitative Analysis}
To analyze the content of AIG-SC guides, we followed a hybrid qualitative coding approach~\cite{azungah2018qualitative} and conducted a thematic analysis over the resulting codings~\cite{braunUsingThematicAnalysis2006}.
To construct the codebook, the team began by hypothesizing codes of interest, including each guide's (1) technical workflow
(2) generation goal,
and (3) computing requirements.
These codes were expanded via open coding to include perceived generation challenges, the use of real identities, and discussions of ethics.
After co-iterating on this codebook through team research discussions and co-coding sessions with multiple team members,
we decided to obtain robust quantitative measures across the dataset for several codes, and thus conducted IRR procedure on a subset of codes we wished to estimate the prevalence over~\cite{mcdonald2019reliability}.
IRR was conducted over a series of rounds between a primary and secondary coder.
In each round, a random selection of 15 AIG-SC guides was independently coded by both coders and Cohen's kappa was calculated for every code.
Afterwards, disagreements were resolved; if necessary, codebook changes were made, and if high reliability was not achieved for any code that round ($\kappa<0.80$), another 15 random AIG-SC guides were selected, and the process repeated.
After eight rounds, high agreement ($\kappa \geq 0.80$) was achieved for every code.\footnote{We make an exception for one code, which only contained five negative instances, making $\kappa$ impossible to calculate in many rounds; the code was confirmed via coder consensus for every negative instance.}
Any codes altered were recoded across the entire dataset.
Our final codebook is presented in Appendix~\ref{appendix:codebook}.

\subsection{Limitations}
Our study holds several limitations.
\textit{First}, we only analyze articles found on \civitai{}, as it is one of the most popular platforms for AIG-SC. However, none of the methods discussed by guides are platform specific and \civitai{}.\textit{Second}, we filter guides by popularity, and it's possible more problematic ones may not be popular (either due to lack of user engagement, or by being quickly removed from the platform). However, this focuses our analysis on techniques that we believe are more frequently shared across AIG-SC communities and thus more likely to potentially aid in NCII creation.
\textit{Third}, we did not analyze guides that only contained attachments, which were frequently usable templates of the workflows explained in guides. While we believe attachments may provide additional insight into how workflows are configured beyond what was visible from the text/images alone, this approach allowed us to focus on what specific details guide creators were focused on teaching or emphasizing to readers.
\textit{Fourth}, we only analyzed guides written in English (the primary language of the paper authors), excluding a subset of guides from our dataset. However, we did identify a few guides in our dataset that were bilingual and do not find evidence that guides in other languages substantially differ.
\textit{Fifth}, while we only consider one form of NCII, NCII does not necessarily just mean sexual content. It is highly contextual and can include images of clothed individuals holding hands or semi-nude content~\cite{chaynExplicitHarmsNonExplicit2026, batoolExpandingConceptsNonConsensual2024}. Future work should expand to include evaluation of such workflows.
\textit{Sixth}, we emphasize that we do not claim to comprehensively capture all workflows, specifically those which are overtly for NCII/CSAM; however, as we find in our results many of these workflows are either the same or substantially overlap, such that these methods can be used effectively there as well.

\section{NCII-Relevant AIG-SC Goals (\ref{rq:aig-ncii})}
\label{sec:ncii-goals}
As stated in the U.S. legislation~\cite{tedcruzS146TAKEIT2025},
NCII is defined as the visual depiction of
(1) a real person's identity,
(2) in ``intimate'' content~\cite{code6851}.
To that end, we analyzed public (accessible to non-verified adult accounts)
AIG-SC guides to determine how often such guides attempt to depict
(1) the identity of a publicly known persona (real or fake), as well as any personas broadly
or (2) `intimate content'~\cite{code6851}, or any sexual content broadly (Section \ref{sec:terminology}).
Through this, we find that, while AIG-SC guides that explicitly focus on AIG-NCII are rare,
many teach readers how to recreate fictional or real identities (\identityGuidesCount), depict intimate (\intimateGuidesCount) or sexually-relevant (\sexuallyRelevantGuidesCount) content,
or discuss how to combine (primarily fictional) identities and sexual content (\intimatexidentityGuidesCount).

\header{Depicting Personas \& Identity}
\label{dataset:persona-content}
The overwhelming majority (\personaGuidesCount) of guides focus on generating, customizing, and training on human (or human-like) personas.
Of these, a substantial number (\identityGuidesCount) attempt to replicate the public identity of a known persona.
While the vast majority (\personaGuidesFictionalCount) are animated, fictional, pop-culture characters (such as animated female characters from TV shows, movies, and video games),
we also find that \realIdentityGuidesCount guides include the identity of real, human individuals, discussed further below.
Although the number of real identities is much smaller than fictitious ones, (as we find in Section~\ref{subsec:architecture}) these methods are seemingly identical.
Many persona-focused guides (\photorealisticGuidesCount) discuss how to create photorealistic personas, including details about skin texture, \gquote{accurate and realistic depictions of human anatomy} (G84) for a diversity of \gquote{body shapes, builds, proportions, heights, and other physical attributes}~(G252).
Non-photorealistic content (\nonphotorealisticGuidesCount) typically focus on anime, caricatures, or anthropomorphized personas.
Guides note that these generated personas can be used for a variety of purposes including commercialization via \textit{``Doujinshi\footnote{Japanese self-published print content (e.g.,magazines, manga).} creation''} (G93),
and \textit{``insta reels''} (G3), or to \textit{``fulfill your dirty fantasies''} (G170). This content may be implicated in ``AI Pimping,'' the practice of placing AI-generated faces onto real bodies and ``using the altered videos to promote paid subscriptions to the AI-generated model’s accounts on OnlyFans competitor sites''~\cite{maiberg24_ai_pimping}, which can both be a form of NCII and labor theft~\cite{europeansexworkersrightsallianceedeepfakeswork26, butlerSexWorkerHuman2025}.

\header{Depicting Intimate \& Sexually-Relevant Concepts}
\label{dataset:sexual-content}
Following our definitions of sexual content (Section~\ref{sec:terminology}), we find that while a small number of collected guides only have sexually-relevant content (\sexuallyRelevantGuidesCount),
or little to no sexual relevance (\nonsexualGuidesCount),
the majority (\intimateGuidesCount) of R+ AIG-SC guides include content that meets the legal U.S. criteria for ``intimate''~\cite{takeItDown}.
This content includes discussions of generating detailed depictions of genitalia and the performance of specific sexual acts.

As AI-generated sexual content can be created for non-abusive reasons (e.g., for sexual expression, creative expression, technical interest)~\cite{minkUnlimitedRealmExploration2026},
advice found within \civitai{} guides supports users in this effort to generate custom content via finer-grained controls:
\gquote{the ultimate goal of AI-driven art is for individuals to customize AI models according to their own aesthetic and creative vision} (G268).
As noted in G269, this can include content that ranges from \gquote{pretty tame} to \gquote{more extreme fetishes.}
Guides claim to teach creators how they can have granular control over what they generate, and train models to understand
\gquote{your character's essence and can place them in a multitude of scenes, outfits, and moods}~(G187).
This can include content that situates personas in a specific setting (e.g., a castle) and/or sexual position. Guides also teach creators how to generate specific objects, such as clothing (\textit{``torn skirt, torn sleeves''} (G190)) or sex toys (G172, G73), that can be composed with a given identity.

\header{Composing Identity \& Sexual Concepts}
AIG-SC guides discuss making \emph{sexual} imagery of a given \emph{identity}.
Of the \realIdentityGuidesCount guides that include the identities of real individuals,
only one discusses how to make intimate content of a specific identity (i.e. AIG-NCII), of an actress;
G27 states
\gquote{While I think its perfectly valid to be attracted to whichever (real life) person, its been a long standing rule to not use them for any NSFW content on CivitAI... this said you can:...}
and lists three methods for creating NCII, noting
\gquote{2. might be inaproriate [under \civitai{} NCII policy] and due to change, so I would recommend 1. for real people, and 3. for fictional characters represented by an actor} (as those methods evade detection; see \iref{irisk:prompting-persona}).
They further note that they aim to make \gquote{a depiction of the character [as] its movie-style, not a depiction of the actress ; allthough I do like the actress.}

Of the remaining three guides, two discuss generating non-sexual content of a real identity: the first uses an image of an actor within a movie as input (but does not describe the output) and
the second discusses how to generate a video of a real identity waving (G28).
Lastly, G261 separately describes how to train, and generate three concepts:
(1) \pquote{a real person's face [and] style}, including \pquote{unique details (e.g. tattoos, intricate accessories),} describing the process as \pquote{Training a Realistic Persona}; the guide also includes an attachment with the name of a well-known female actress.
(2) \pquote{specific dresses, uniforms, or costumes} such as a \pquote{bikini, [or] tank top,} noting how to \pquote{avoid tying [the outfit] to one face} for application onto any identity.
(3) \gquote{NSFW/Adult} concepts including \pquote{masturbation, [female genitalia]... general sex scenes.}
While these concepts are discussed separately,
concept composition is taught in several other guides (G197, G269).

Most guides that discuss how to make intimate content of identities (\intimatexidentityGuidesCount) focus on publicly known fictional identities, such as
Elsa (from the animated film \textit{Frozen}) and Tifa Lockheart (from the video game \textit{Final Fantasy}).
Beyond applying intimate concepts to specific identities, several guides also discuss how to apply them to \emph{arbitrary} identities.
These guides may discuss \textit{``How to generate NSFW [fellatio] videos with any image''}~(G100), or note that they \gquote{rarely use real people in my creations} but specify that \gquote{for real people, I'd suggest just sticking to these two [model ``adapters''] for the best results} (G275).
Although few specify whose identities could be portrayed, many brazenly hint, or explicitly state, that these may include real identities.

\section{What Architecture Supports AIG-SC, and How NCII-Risks Are Enabled (\ref{rq:architecture-and-risks})} \label{subsec:architecture}
By analyzing technical descriptions of \numGuidestotal guides, we identify the core components of AIG-SC workflows, develop a framework to describe the processes, data, and models used in the production of AIG-SC workflows,
and model how existing guides introduce AIG-NCII-relevant risks.

\header{AIG-SC Workflows: Basic and Advanced}
To better model more simplistic and more involved attackers,
we define two AIG-SC workflow variants.
\textit{Basic workflows} consist of the minimum set of components required to generate AIG-SC: a single prompt, a single `base model,' which processes the prompt, and their resulting output.
\textit{Advanced workflows} consist of any workflow that, in addition to a basic workflow,
(a) uses another model to `adapt' the base model,
(b) trains a base, or `adapter' model, or
(c) links multiple AIG-SC workflows together (e.g., one workflow's output serving as another workflow's input).

While we use these terms to denote their relative technical involvement, we find and emphasize that these workflows are broadly usable by laypersons. This is in large part due to reusable, GUI-based tools and templates used throughout guides: \gquote{these pictures and videos can be dragged and dropped into the Comfyui [generation tool] interface allowing you to see the workflow/metadata that generated that particular image or video} (G103). Further, we find evidence that skilled creators are translating technical advice from papers hosted on arxiv.org to \civitai{} users. For example,
one guide offers an extensive comparison between two optimization algorithms used for fine-tuning, one of which was first published in ICML (International Conference on Machine Learning). As the same guide describes, \gquote{savvy practitioners tell noobs to read the white papers, but the white papers have too many words, loss graphs and math}~(G279), which led them to create their guide with detailed results on how the optimization algorithms impact training.

As a result, many guides note that despite not having a technical background, they are now able to effectively create complex AIG-SC workflows:
\gquote{I don't have any formal training or background in this -- I just really got into the topic and started experimenting} (G263).

\header{Threat Model}
To gain a first understanding of ``reasonably foreseeable misuse''~\cite{euAIAct} of an AI workflow for AIG-NCII,  we consider how a single adversary may use a conventional image or video AI generation workflow to produce AIG-NCII which, by definition~\cite{tedcruzS146TAKEIT2025} must contain
(a) an intimate depiction
(b) of a real identity.
As concepts within AIG systems are easily composable once introduced~\cite{liu2022compositional},
we analyze AIG-NCII risks by separately considering how
sexual (\sexualrisklabel)
and identity (\identityrisklabel) concepts are introduced or elicited in AIG-SC workflows.
Without loss of generality,
we consider that the adversary is targeting the identity of a real person with a digital footprint containing a reasonably small set of videos and images (either from the public internet or semi-private social media accounts)-- in cases where specific identities (such as public figures) may be uniquely at risk, we highlight in-text.

The adversary can freely modify or extend the data, models, or processes used in conventional AIG workflows.
The adversary does not directly have access to any existing NCII of the identity (either via data or models), but does have access to the public internet and an open-weight AIG-SC ecosystem that (comparable to \civitai) attempts to remove models representing specific identities (intimately or otherwise).
For the purposes of this paper's scope, we abstract away any specific software and/or hardware required to run this workflow, and only highlight when data, model, and computational availability/costs are uncommon/higher than those readily found in existing open-weight AIG ecosystems, or beyond the capabilities of conventional consumer hardware (e.g., entry-level laptops).
We re-emphasize these capabilities do not require niche skills or a technical background (as stated by the guides themselves) and are within reach of many motivated attackers.

\subsection{Basic Workflow}
\label{sec:basic-pipeline}
\begin{figure*}
    \centering
    \includegraphics[width=\linewidth]{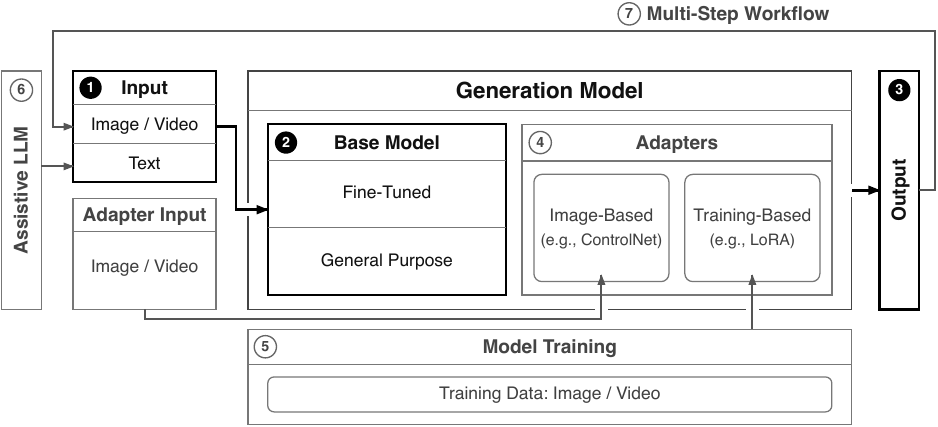}
    \caption{\textmd{\textbf{Architecture of AIG-SC Workflow} --- {\small The technical components of an AIG-SC workflow; optional \emph{advanced workflow} components depicted in grey.}}}
    \label{fig:aig-sc-pipeline}
\end{figure*}

As shown in the bold components of Figure~\ref{fig:aig-sc-pipeline}, a basic workflow consists of
\ding{182} a text input,
\ding{183} a single base model,
and result in \ding{184} an output.
All AIG workflows either use, modify, or extend this basic workflow.

\header{Base Model (\ding{183})}
All generation workflows require an existing base model to translate inputs into outputs; \modelsGuidesCount guides mention one or more base models by name. Guides suggest that creators choose models that are best suited, often as a result of their training data, for generating certain types of content: \gquote{SDXL is trained to make skin look real}~(G256).
As AI-generated outputs on \civitai{} often detail which model was used, creators may  look for \textit{``images that make you go Woohoo!''} (G170) to identify models capable of generating desired content. Nearly all
of the models mentioned by creators are widely accessible on \civitai{} or other public repositories.

Many guides use general-purpose models (\generalModelsGuidesCount), including Stable Diffusion models (SD1.5, Stable Diffusion XL~\cite{podell2024sdxl}) or WAN~\cite{wan2025wan}, to generate output media. All but four of these guides mention at least one open-weight general-purpose model, which allow other developers to modify the model weights (fine-tuning), and subsequently change the output behavior of the model; as we find next, fine-tuned models are often used in the guides. The result of large-scale training efforts, general-purpose models are capable of producing a diversity of content: \textit{``after two years, SDXL is the most trained image generation model out there...it's wildly creative and arguably the king of photorealism...it's versatility and quality make it a go-to for image generation projects''} (G262).

Many guides (\finetunedGuidesCount) also describe workflows that use \emph{fine-tuned} models, open-weight general-purpose models that have been further trained to generate specific content more accurately (Section~\ref{sec:training}).
Fine-tuned models are frequently used; the most popular model across the guides, Illustrious (Table \ref{tab:all-models}), is a fine-tuned derivative of open-weight model SDXL. Some, like Pony (also fine-tuned from SDXL), may be trained for ``stunning SFW and NSFW visuals of various anthro, feral, or humanoids species and their interactions''~\cite{ponyRef}, while others, like NoobAI (fine-tuned from Illustrious), are trained for \gquote{the broadest built-in knowledge of anime characters and artists}~\cite{noobRef}.
These models
are noted to leverage training data from websites like Danbooru, a popular imageboard that hosts millions of anime images organized by user-generated keyword tags (e.g., 1girl, 1boy, blurry, t-shirt, closed-mouth, masterpiece)~\cite{danbooruRef}.

\ispara[irisk:models-with-embedded-peronas][srisk:models-with-embedded-sexual-concepts]{General-Purpose Models are Already Trained on Real Identities}{and Intimate Concepts}
As general-purpose models are trained on billions of inputs~\cite{schuhmann2022laion},
prior work has shown that they contain sufficient representation
to generate the likeness of public personas within that data~\cite{yuan2023inserting};
of the \identityGuidesCount guides that discuss generating a public identity, six use general-purpose models, and
some model pages on \civitai{} even include images of celebrities as example model outputs~\cite{ortegaAILovesBoobies2026}.
They are also capable of generating sexual or sexually-relevant concepts~\cite{li2024safegen}; we emphasize that these outputs are not inherently abusive when not paired with identifiable personas. We find guides that describe using general-purpose models like SDXL and Flux to generate \gquote{popular types of bras and panties} (G259) and personas with \gquote{deep cleavage visible, missionary pose...boobs bouncing} (G252) without any mention of jailbreaking.
However, some guides do make jailbreaking suggestions: \gquote{I typically use `Moves seductively' as a safe way of having sexy movement...can still be a little hit or miss but it's a safe start. But can sneak in a `breasts bounce' typically without a problem}~(G86).

\spara[srisk:sexual-intended-model]{Fine-Tuned Models (Re)introduce Sexual Content}
\label{subsub:models-sexual-modifications}
While some general-purpose models, like Stable Diffusion XL, use data-filtering procedures to remove sexual data~\cite{StableSafety}
a number of accessible fine-tuned models appear to be explicitly retrained on sexual content, as made clear by model names such as \gquote{CinEro | Pony | NSFW v0.9d} (G277), \gquote{Big Love Photo1} (G262), and
\gquote{WAI-NSFW-illustrious-SDXL} (G91), in which intimate content is generated even without being specifically prompted and in fact must be explicitly prompted against:~\gquote{users are expected to consciously add `nsfw' to negative prompts to [prevent the generation of] inappropriate content} (from the linked model page on \civitai).
Similarly, G90 notes that they include \gquote{manly} in prompts \gquote{to reduce the yuri\footnote{\label{fn:yuri}A Japanese genre of lesbian fiction.} influence} of their model.

\header{Text Inputs (\ding{182})}
\label{subsec:text-prompts}
\civitai{} guide creators use text prompts (\textInputGuidesCount) to instruct the generation model on what should (positive prompts) and should not (negative prompts) be generated. Text prompts are typically constructed from a combination of natural language and/or model-specific keywords.
In crafting text prompts, guides often mention the difficulty of dealing with ambiguity in language and the resulting output:
\gquote{what one person calls a squat, another sees as a crouch and so on} (G72).
As a result, guides suggest that prompts have \textit{``clarity and specificity''} (G84).

\spara[srisk:sexual-prompting]{Craft Anatomically Accurate Sexual Prompts}
As most workflows rely on open-weight models, most workflows are not affected by safety filters, and to include realistic sexual content, simply recommend \gquote{using correct and specific anatomical terms} which  \gquote{significantly improves the model's ability to render accurate depictions. For example, terms like `labia majora', `labia minora', `clitoral hood', `vulva', `erect penis' and `vagina' provide clear guidance} (G84).
Several guides also use LLMs and custom-built tools to assist in prompt creation (\sref{srisk:sexual-prompts-via-llm}).

\ipara[irisk:prompting-persona]{Depicting a Public Identity (or Anyone) Without Their Name}
To replicate the likeness of public figures via prompting (which we term \textit{likeness prompting}),
several guides mention that to get around the \gquote{long standing rule to not use [the name of real public identities] for any NSFW content on CivitAI}, they instruct others to \gquote{refer to the movie/series name and the fictional character's name ... without using their [real] name, so they AI wont use their exact image, but at least get its inspiraration by the attriutes}~(G27).
Furthermore, the same guide also notes that if that is not allowed, one can \gquote{use any LLM (ChatGPT, DeepSeek, Gemini, whatever) to describe the person and/or character they portrait, that you like, and used that description instead of their names}, such that no direct identifiers to any identity (real or fictional) are used.
This same method of image-to-text prompt construction is also mentioned by custom software tools to enable the generation of any text-based prompting of any identity, with an existing image~(\iref{irisk:identity-prompts-via-llm}).

\header{Image Inputs (\ding{182})}
\label{subsec:image-inputs}
Some creators (\imageInputGuidesCount) start from an input image to modify that image or use it as a starting point for video generation.
Workflows use inpainting techniques, which require an additional step, in order to modify inputs with more precise control (Section \ref{subsec:multi-step}).

\ispara[irisk:nudification-identity][srisk:nudification-sexual]{Provide an Image of an Identity}{and Prompt for Nudity or Sexual Content (e.g., Undressing)}
Creators can use a sexual text input (\sref{srisk:sexual-prompting}) and image input depicting an identity, to generate sexual content of that identity; when this identity is real, this directly generates NCII.
For this reason, \civitai{} disallows all nude-manipulating workflows. However, our dataset still captures evidence of one.
G197 describes how to make an image-to-video workflow of a sexual act.
They further link to a \gquote{universal workflow} page that describes how to use this workflow more generally.
Across two versions of the description (also posted on \civitai), they state that generation is guided by a \gquote{one-line-scenario} text prompt; and, as a syntactic example, notes that a \gquote{Good idea for prompt[s are]: `Girl danse and undress', `Girl flirting and smile poses on camera,'} illustrating that undressing is as simple as any other text prompt. Later, undressing is used as another technical exemplar \gquote{undress wooman and take off sweater}-- the guide never addresses where the image input to the workflow should come from.
Notably, beyond these text prompts, the guide otherwise conforms to \civitai's policies, with no technical details intrinsically tied to undressing--this illuminates an important point that recurs throughout our findings; the difference between AIG-SC and AIG-NCII is not the vast, legal technical infrastructure that powers AIG-SC, but minor, extremely accessible changes within these workflows.

\subsection{Adapting Base Models}
\label{subsec:adapting-model}

\begin{figure}[t]
    \centering
    \includegraphics[width=.65\linewidth]{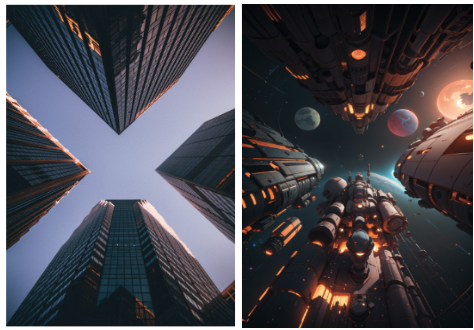}
    \vspace{.5pt}
    \caption{\textmd{\textbf{Image-Based Adapter Example} --- {\small ControlNet~\cite{zhang2023adding} takes an input image (left) to generate a new output that resembles the input \textit{structure} (right)}}~\cite{civitaiControlNet}.}
    \label{fig:control-net-example}
\end{figure}

As shown in Figure~\ref{fig:aig-sc-pipeline}, several guides (\adaptersGuidesCount) discuss advanced workflows that augment the basic workflow by \textit{adapting}  the base model.
These `model adapters' (most notably, LoRAs~\cite{hu2021lora}) provide lightweight modifications to base models without additional training. They incorporate one or more concepts learned from images from a (small) set of training data, including sexual content or identities that may be hard to prompt for, or have been intentionally excluded from base models~\cite{ortegaAILovesBoobies2026,cretuEvaluatingConceptFiltering2026}.

\header{Image-Based Adapters (\ding{195})}
\label{subsec:image-adapters}
Unlike prompting which
\gquote{requires careful attention to language and detail[,]... Iterative refinement and adjustment of prompts[,] ... [and] remaining patient and persistent}~(G84)
or pre-trained adapters, which require either that a given adapter is available or that one has enough resources to train themselves (Section~\ref{sec:training}),
image-based adapters are already trained to extract the high-level structure from any provided image.

Given an input image, image-based adapters like ControlNet~\cite{zhang2023adding} can constrain the resulting generation (Figure \ref{fig:control-net-example}) to ensure \gquote{the underlying structure [of the adapter-input image] remains perfectly consistent} in the output (G29) which \gquote{can help to reduce the number of assumptions the model has to make when just using a prompt} (G92).
For this reason, several guides (\imageAdaptersGuidesCounts) use image-based adapters. Within persona-focused images, G92 notes that this allows \gquote{the model [to] have a better idea of the number of subjects} to generate in an output image
and ensures that the structure is maintained
\gquote{no matter how radically we change the subject, lighting, or setting}~(G29).
Similarly, guides also use  IP-Adapter~\cite{ye2023ip}, another image-based adapter, to
\gquote{copy the shape, face, color, style, etc of the input image... [which helps ensure] distinct facial features}~(G16) from the adapter-input are replicated in the output.

\spara[srisk:sexual-image-adapter-inputs]{Transfer Sexual Poses or Situations from an Input}
Several guides note that using tools like ControlNet ensures that generated images accurately depict a desired sexual situation, such as \gquote{yuri\textsuperscript{\ref{fn:yuri}}scenes, threesomes or group sex images} (G269),
if also depicted in the image provided to the adapter.
Guides note \gquote{you can source these... images from anywhere AI-generated [images],
high-resolution stock photos and the like} (G29) and, if you \gquote{can't find an image that depicts the weird things you want your characters to do...use GIMP and make one yourself!} (G92).
This structure frees creators to have more granular control between who is depicted and what they are doing
\gquote{because [ControlNet] frees us from the base image's specific details. We are no longer copying a[n] individual person; we are mimicking a pose, a perspective, and a spatial layout} (G29).

\header{Pre-Trained Adapters (\ding{195})}
\label{subsec:pre-trained-adapters}
Several guides (\preTrainedAdaptersGuidesCounts) modify base models with pre-trained adapters,
which can generate outputs that depict concepts learned by training on an image dataset containing that concept.
While there are a variety of adapter methods,
low-rank adaptation~\cite{hu2021lora} (LoRAs) are most common across the guides.
As LoRAs have small file sizes (e.g., 35MB), they can be easily shared across consumer laptops and platforms (\gquote{there are many Loras in Civitai for making videos} (G2)),
and require few resources to use and train (\sref{srisk:cheap-training}/\iref{irisk:cheap-training}).

\ipara[irisk:loras-of-people]{Guides for Replicating Real or Fictitious Identities Use the Same Methods}
\label{text-only:lora-selection-personas}
Pre-trained adapters, particularly LoRAs, can be used to generate consistent outputs of real identities (notably of public figures)~\cite{hawkinsDeepfakesDemandRise2025}, and were previously hosted en masse on \civitai{} until a recent payment processor pressured policy changes~\cite{maiberg25_civitai_people_ban}.
We did find several workflows (\preTrainedAdaptersIdentityGuidesCounts) that contain a LoRA whose \gquote{purpose is to produce a specific character} (G79).
LoRAs enable accurate replication that would otherwise be difficult to achieve: \gquote{Character LoRAs help you generate a particular character that cannot be generated with a[n existing model]} (G90).
Whether the character is a real identity or not, the generation process is the same: by \gquote{insert[ing] the prompt given by the character LORA} (G89) into the text input, one can recreate any \gquote{real person's face [and] style} (G261).
Further, some adapters appear to be trained from datasets that include the identities of real people such as \gquote{InstaGirl} (G197), whose corresponding website describes it as \gquote{the result of a meticulous training process on a diversified dataset of 111 curated high-quality images;} we find this practice is encouraged within training guides (\sref{srisk:data-sexual-content}/\iref{irisk:data-identifiable-people}).

\spara[srisk:lora-sexual-modifications]{Adapters Are Used to Apply Sexual Concepts}
\label{text-only:lora-selection-sexual-modifications}
While many models used across guides can already generate sexual content (\sref{srisk:sexual-intended-model}) and one can try to accurately prompt for every anatomical detail (\sref{srisk:sexual-prompting}), guides mention that
adapters are extremely helpful in situations \gquote{where `highly customizable' comes into play. Use specific LoRAs to tweak the output... [in ways] that prompting alone cannot achieve} (G29).
Many guides (\preTrainedAdaptersSexualGuidesCounts) discuss using pre-trained adapters for sexual content, which apply to persona concepts including the use of sex toys or placement of body parts (e.g., hands, legs)
during sexual acts.
These LoRAs are also used to specify details about how personas should look; for instance, to enable precise control over anatomy like \textit{breast size} (G273), make subjects \gquote{completely nude} with visible \gquote{male pubic hair} (G10), or broadly make \gquote{sex movements and animations a lot smoother}~(G269).
AIG workflows can include multiple LoRAs for a single generation; for instance, G269 includes \gquote{three of four LoRAs}, specifying a \gquote{general NSFW LoRA, some kind of bouncing breasts LoRA, and a LORA that specializes in whichever activity is most prominent in the video.}

As noted by guide~G29, LoRA-based composability is not limited to sexual content:
\gquote{Do you want a specific body type, a particular style of photography, [and/]or a certain aesthetic filter? Stacking the right LoRAs on top of your base model allows for granular control}~(G29). When combined with \iref{irisk:loras-of-people}, multiple LoRAs can directly facilitate AIG-NCII through the composition of identifiable, intimate concepts.

\subsection{Training (Base/Adapter) Models}
\label{sec:training}

Several guides (\trainingGuidesCounts) discuss how to train models (\ding{196}) to generate content not \textit{``reproducible in the base model simply by prompting''} (G168).
During training, models update their weights to learn patterns between inputs and outputs of the dataset---in our guides, this learned pattern is often between images and their corresponding text captions.
If training datasets teach models how to generate sexual content or identities not previously captured, it can enable AIG-NCII, as \gquote{your LoRA will only ever be as good as the images you train it on} (G187). Subsequently, the choice of training data has a direct impact on AIG-NCII.

\sipara[srisk:data-sexual-content][irisk:data-identifiable-people]{Training Datasets for Sexual Content}{Contain the Identities of Real (Non-Consenting) Individuals}
Guides suggest creators source data from \textit{``porn sites like reddit, twitter, ashemaletube, bdsmir''} (G175); no guides mention obtaining consent from those depicted in online images.
As the consensual sharing of an individual's sexual image is not consent for new sexual content~\cite{geengUsableSexurityStudying2020, qinDidTheyFing2024}, intimate outputs of these identities are NCII.
Furthermore, as NCII is distributed on sites such as Reddit~\cite{conversation2015_reddit_revenge_porn, huberNonconsensualIntimateImage2024}, any large-scale data collection of these sites is likely to perpetuate the harm of existing victim-survivors.

Guides emphasize that creators do not need to be limited to traditional pornography, and instruct readers that they \gquote{can search for images absolutely anywhere. There are no restrictions} (G167) and direct them to \gquote{grab [images] from the web or generate with AI tools (even Grok can help!)} (G261).
As adapters trivially facilitate the combination of identities and sexual concepts (\sref{srisk:lora-sexual-modifications}/\iref{irisk:loras-of-people}), several guides could be interpreted as enabling the replication of others across the internet, and the NCII they enable.

Other guides simply provide large-scale datasets of sexual content of real identities directly. For instance, G234 boasts of providing \gquote{the largest program for searching and downloading sex images... with its caption, ready for training.}
They do not discuss how the images were obtained, but clarify:
\begin{quote}
\gquote{All the attached images are AI-generated because \civitai{} policies do not allow the publication of real people's images. However, the program contains over half a million real images that can be used for LoRA training.}
\end{quote}
We discuss how guides view the ethics of their actions, as well as interpret and interact with \civitai{} policies in Section~\ref{sec:boundaries}.

\sipara[srisk:cheap-training][irisk:cheap-training]{Training Models on Sexual Content}{or Identities is Accessible and Cheap}
The availability of well-curated data, software, cloud-based hardware, and peer-advice through guides is frequently noted as having significantly lowered the barrier to even custom model training and creation:
\pquote{I always thought the whole process of LoRA creation to be very technical and difficult, but I realized I'm very wrong,} noting that the platform significantly helps, with \gquote{most of the heavy lifting done by CivitAI}~(G271).

Due to breakthroughs in low-resource training, such as LoRA~\cite{hu2021lora}, the data, hardware and software requirements of training workflows are increasingly within reach of consumers, with one guide reporting that the \gquote{total training time was roughly 4 hours} for a LoRA with a consumer-grade\gquote{RTX 5090 [Graphics Card],} or by using cloud GPU services that cost \gquote{\$0.9 an hour} (G271).
Data requirements have also plummeted; validating prior findings, what previously required thousands of images~\cite{minkUnlimitedRealmExploration2026, hawkinsDeepfakesDemandRise2025, han2025characterizing}, now just needs \gquote{a good starting dataset of 25-40 images} that is \gquote{high-resolution (1024x1024 or higher), clear, and well-lit} (G187).
To further lower data-based barriers, guides such as G176 encourage using \gquote{data augmentation,} noting that even in small datasets \gquote{perspective diversity matters in LoRA... [and] a [certain perspective] may only have 5-10 images, which might not be enough,} but that one can apply \gquote{simple augmentation techniques like rotating, flipping, or scaling the images to artificially increase their variety... without needing more original images.}

As a result of this accessibility, G271 notes that costly resources, knowledge, or time is not needed --- even when they were \gquote{a complete beginner, I didn't touch any settings at all and the LoRA still turned out fine,} and goes on to note that with this access, creators are no longer limited to platform-curated content: \pquote{Ever since I figured out how easy the process was...it's just repeating it... and you can now create the characters that you like but aren't on the platform}~(G271).

\subsection{Multi-Step Workflows}
\label{subsec:multi-step}
A number of guides use \emph{multi-step} workflows that employ more than one AIG model and generate intermediate outputs (e.g., text, images) during AIG-SC creation.
Importantly,
multi-step workflows can accelerate novice development,
translate content between mediums,
and provide more granular control over identity generation, which has dual uses for AIG-NCII.

\header{Assistive LLMs (\ding{197})}
A number of guides (\llmGuidesCount) use Large Language Models (LLMs) to assist in AIG-SC generation, often aiding in setup, conveying knowledge, or crafting inputs.

\sipara[srisk:llm-upskilling][irisk:llm-upskilling]{LLM Assistants Lower the Barrier of Content Generation for Sexual Risks}{and Identity Risks}
Several guides mention that production LLMs are essential for teaching readers how to set up and run AIG-SC workflows locally.
After discussing how to generate a sexualized depiction of two fictional identities, G92 concludes by noting that
\gquote{If you're new to open source code or can't get an extension to work - just ask around! (not always me, specifically - there are AI chat bots that'll walk you through the process).}
However, this goes beyond just software installations;
G172 notes that frontier LLMs are incredibly helpful for generalized learning about local AIG-SC generation, noting
\gquote{the best resource to learn to train is probably ChatGPT. The AI model knows a lot about AI models, can scan GitHub repos, etc... This makes it an excellent consultant for this kind of work. I used it extensively while I was learning.}
Furthermore,
LLMs also enable local generation by assisting creators in modifying and configuring AIG-SC workflows to suit their unique hardware and software constraints.
G2 notes that \gquote{after a lot of headaches, I managed to make videos,} and mentioned that
\gquote{I wrote this [prompt] to Grok and [it] helped me out tremendously:}
\gquote{Hi Grok! I'd like you to help me install ComfyUI [generation software] on my laptop. I have an RTX 4090, Windows 11, an Intel i9, and 900GB of hard drive space. I'd like to be able to use the Wan video and Anisora models in local mode. Can you help me?}, noting to readers: \gquote{(You change the message according to your PC specifications.)}

\ispara[irisk:identity-prompts-via-llm][srisk:sexual-prompts-via-llm]{Turning an Image of an Identity into a}{Sexualized Text Prompt}
G24 uses an LLM to generate a sequence of text-inputs that describe a specific identity (provided via an input image) in an increasingly sexual scenario. First, they generate an image of a physical background (\gquote{Swimming Pool Changing Room}) and of a specific persona (in clothing). Using these two images, they ask ChatGPT to \gquote{create a new prompt that places the woman in...the room.} In a series of exchanges with ChatGPT, using jailbreaking technique multi-turn prompting~\cite{song2026multibreak, yang2025many}, they generate a sequence of prompts that describe that input persona in an increasingly sexual way: \gquote{the first 2 prompts should be fully clothed and safe-for-work. The next 2 prompts should be artistic nude or implied eroticism. The last 3 prompts should be explicit, very erotic NSFW with intimate close-up details.} These text prompts could then be provided to an image generation model; if the input image contains a real identity, this could facilitate the generation of AIG-NCII.

\header{Inpainting (\ding{198})}
Rather than providing an entire image as a direct prompt to the basic workflow, inpainting workflows specify a subarea of an image to be modified. After identifying a section to alter (via a ``mask''), creators then specify how it should be altered, either with a text-input or with a different input image.
This workflow is especially useful for correcting small imperfections; for example, G79 corrects \gquote{a hand [that] has missing fingers} by replacing it with an image of another hand, something that, with text prompting alone, \gquote{would require the proverbial thousand words in specificity.}

While some guides have creators outline an image's masks manually, others use ``detailers'' --- a method for automatically detecting an image region (via models like YOLO~\cite{redmon2016you}) and inpainting that region, which is often used to fix commonly noted issues in hands, eyes, or breasts of images.
We find that this technology can also be used for face-swapping.

\sipara[srisk:faceswapping-sexual][irisk:faceswapping-identity]{Provide an Intimate Image}{and Apply an Identity (e.g. Faceswapping)}
Guides discuss using automatic masking tools to swap the face from one input image onto the face of a persona in another, sexual image, a workflow that mirrors those supported by ``dual-use'' faceswapping applications~\cite{daffalla2026dual}. One guide shares a one-line jailbreaking technique for automatic face masking tool ReActor that will \gquote{do the trick} so that those originally \gquote{unable to output an image with either a spicy input or source image} (G17) can swap a new face onto sexual content, directly enabling the generation of AIG-NCII.

However, guides take this a step further, proposing multi-step workflows that first generate customized sexual content depicting unknown personas and then modifying the identities of the personas in that scene, allowing them to follow a faceswapping workflow without needing to first source sexual content. For example, G8 first generates a sexual content image via text-inputs, with keywords \gquote{masterpiece}, \gquote{2girls}, \gquote{femdom}, and \gquote{lifting, imminent kiss.} Then, it creates a mask for one of the personas in the scene and uses a LoRA to ``swap'' the identity of the unknown persona with a generated depiction of a known (fictional) persona. This example also does not require an input image of the target---simply an identity LoRA (\iref{irisk:loras-of-people})---and allows attackers to further customize the nature of the sexual content without inputs.

\header{Generating Photorealism Through Anime Models (\ding{198})}
G157 presents a multi-step workflow to generate high-quality, photorealistic outputs using anime models. The guide uses a \gquote{well-composed anime-style} image of a persona as an input
to a \gquote{realistic model (Illustrious/Pony hybrid).} The guide further uses image-based adapter ControlNet which \gquote{ensures structural consistency} is extracted from the anime image input. This workflow generates \gquote{a photorealistic version of your input image with enhanced detail and quality} that the creator finds \gquote{leverages the complementary strengths of different AI models to achieve superior results.}

\ipara[irisk:style-modifications]{Modifying Anime Content to be Photorealistic}
As this workflow demonstrates, even anime models can support AIG-NCII creation; this workflow allows creators to transform sexualized depictions of  anime personas, which can still depict real identities, into photorealistic sexualized outputs. Prior work has found that anime-style content is one way that people justify AIG-NCII~\cite{minkUnlimitedRealmExploration2026}. Further, as discussed in Appendix \ref{app:age-risks}, the use of anime-style models may increase the risk of generating younger-aged personas which, if photorealistic, may constitute CSAM (\aref{arisk:style-decreases-age}).

\section{Ethics and Governance (\ref{rq:aig-sc-guides-ethics})}
\label{sec:boundaries}

While some creators use the previously described workflows to generate content depicting
entirely fictional characters, as demonstrated, they also enable the generation of AIG-NCII.
As such, the establishment of  norms, rules, and their enforcement is essential for restricting discouraged actions, such as AIG-NCII~\cite{qinThoughtYouWere2026}.
However, we find that, while \civitai{} creators are somewhat concerned with legal compliance and fair use, discussions around ethical boundaries are infrequent and, in some cases, explicitly focused on evading platform restrictions.

\header{Legal Compliance and Fair Use}
Guides sparingly reference legal restrictions on content creation. However the language used suggests that they are primarily trying to cover a legal basis rather than a genuine desire to establish a boundary, which may ultimately act as a baseline for community-based governance~\cite{lapointeGovernanceAIgeneratedPornography2026, qinThoughtYouWere2026}.
For example,
G15 posted a jailbroken model with text clarifying it was  \textit{``published 1st August 2025 (if the EU is watching),''} likely referring to the EU AI act which requires that anyone releasing new or modified models (as of August 1, 2025) follow certain transparency regulations~\cite{euAIAct}. One AIG-NCII enabling guide, that teaches readers how to use a faceswapping tool \gquote{with either a spicy input or source image,} ends the guide with, \textit{``for your own good, stay away from the likenesses they tell you no no about''} (G17).

Legal concerns extend to fair use,
particularly when sourcing inputs in relation to art (e.g., illustrations) due to concerns around copyright, \gquote{When working with NSFW content or any sensitive material, it's crucial to adhere to all legal and ethical guidelines, including respecting copyright} (G254).
Since the Illustrious model was trained on anime-style content, the team allows artists to remove their content from training. G205 explains that: \gquote{If any artist does not wish to see their work in Illustrious, they must contact the original creators of these models} (G205). However, this only affects future model training since Illustrious has already been publicly released.

Unsurprisingly, these concerns around consent and fair use did not translate to other forms of data, including data that is non-consensually scraped from social media sites or adult content sites to train models.

\header{Community Governance Against AIG-NCII}
We find that only one creator explicitly names NCII as a possible outcome for following their guide, stating \textit{``please don't use this to make deepfakes of people without their consent. Please be responsible and be a good person''} (G270). While some guides include disclaimers or requests for ethical use,
they are often directly contradicted by other guide content or lack contextual details. For example, one guide describes how to prompt for sexual content of public figures without using their name (\iref{irisk:prompting-persona}) and then notes that, given the \gquote{long standing rule to not use them [real life people] for any NSFW content on CivitAI}, this \gquote{might be [sic] inaproriate} ; as Ortega et al. similarly finds, the actual intention behind statements like ``don't make children'' are unclear~\cite{ortegaAILovesBoobies2026}, but may be motivated by fears of platform banning \textit{``...if u using photos of ppl to train loras on civitai and you can see their faces u can get banned because civitai doesn't allow you to train on ppl's faces''} (G175).

\header{(Evading) Platform Policies}
We find evidence of guides discussing how to generate content that they know is disallowed by platform policies. For example, G86 begins: \gquote{Ever seen those breast enlargement videos and thought, `Wait a minute... how did they pull this off when AI video generators shut down any prompt that even whispers the word `breast'?'} and continues to describe how to generate two (non-sexual) input images, and \gquote{add the image with small breasts to first frame and the image with big breasts to second frame and click on generate} to produce sexual content on a video generation site that disallows sexual generation. Similarly, we find a guide that teaches how to use a sexualized text prompt to \gquote{make [their chest] bouncy and NSFW as possible} noting that, for the specific model chosen by the creator (a closed-generation model), \gquote{as long as the image is covered on their bits you should 90\% of the time be safe} (G86) to generate sexual movement. 

\header{Norms and the Broader AIG-SC Ecosystem}
AIG-SC is supported by a broader ecosystem of online communities (e.g., on Reddit, Discord)~\cite{doringExperiencesAIGeneratedPornography2025, medeirosCharacterizingResourceSharing2026, qinThoughtYouWere2026}; some specialize in providing real-time technical support for content creation and assist in translating guide content to implementation~\cite{qinThoughtYouWere2026}. \civitai{} can serve as an entry point and be used by creators on other platforms to advertise their communities. Within guides, we find direct links to communities on WeChat, Discord, and Reddit. While most are general-purpose AIG-SC communities, one of the advertised communities was specifically created to preserve previously deleted abuse-enabling models from \civitai.
We also find solicitations for contact between \civitai{} users: \gquote{if you have any questions, you can leave me a message} (G23), which can include methods of contact that are off-platform.
Within the guides, we found that some creators provided their personal email, Twitter/X handle, Github username, Patreon link, YouTube, and more. This demonstrates that \civitai{} is directly connected to other platforms.
Lack of community-based governance on \civitai{} could lead those seeking to create abusive content off-platform into spaces that do have stated rules against such content.
Particularly when coupled with community norms that may make it uncomfortable for members to vet a creator's intended use case prior to providing assistance~\cite{qinThoughtYouWere2026}, those seeking to create abusive content can seek assistance with multi-step workflows through external communities.
Whether \civitai{} funnels those creating abusive content into external communities and how that may affect them is a subject for future work.

\section{Discussion}
\label{sec:discussion}
From discussion groups within bulletin boards (BBS) in the late 1970s to Usenet sexual interest groups in the 1990s, to modern-day AI-generated sexual content communities, anonymity on the internet has always provided space for people to discuss their sexual interests (some of which may be otherwise stigmatized)~\cite{samanthacoleHowSexChanged2022}.
Accordingly, the challenges of how to
govern online spaces to allow for sexual expression while curbing abuse are not new, nor is the non-consensual creation of intimate content~\cite{mcglynnRevengePornContinuum2017}.
However, unlike the introduction of camera-enabled smartphones, which has greatly exacerbated its creation~\cite{vitisPrivateHiddenObscured2020, chaMyBodyNot2026,henryImageBasedSexualAbuse2023},
AIG-NCII can be created at scale, with ease, and with little contact or data from the victim-survivor.
Furthermore, the technology that powers this abuse is often used by
AIG-SC creators with no intention to create AIG-NCII, as well as
general-purpose image- and video-generation technology.

While prior work identifies this resource overlap~\cite{minkUnlimitedRealmExploration2026, qinThoughtYouWere2026, medeirosCharacterizingResourceSharing2026}, our findings show that
these resources, which are designed to generate highly customized content, are part of real workflows shared by \civitai{} community members and therefore should be tested for the capability to
support AIG-NCII through the composition of both sexual (\sref{srisk:models-with-embedded-sexual-concepts}-\sref{srisk:faceswapping-sexual}) and identifying concepts (\iref{irisk:models-with-embedded-peronas}-\iref{irisk:style-modifications}) across a number of technical components.

\header{NCII Creation is a Workflow Configuration}
AIG-NCII is not enabled by novel technology that can be independently regulated, but by a set of configuration choices (e.g., models, model adapters, inputs) that creators make when setting up otherwise innocuous generation or training workflows. This suggests that many AIG-SC guides on platforms like \civitai{} may, whether intentionally or not, be ``dual-use'' and aid offenders in generating abusive content, similar to identified dual-use applications~\cite{daffalla2026dual}. As noted, we found only one guide that explicitly discussed the process of generating AIG-NCII. However, a subset of guides were written in a way that made clear that, for example, an image-based sexual content generation workflow could be run with a real identity. As we demonstrate in Section~\ref{sec:boundaries}, technical advice on \civitai{} is rarely if ever accompanied by meaningful attempts at setting community-based norms that discourage NCII creation and/or emphasize its harms. Instead, users may be cognizant of \civitai's policies against NCII creation and intentionally modify their word choices to evade detection, such as using the
word ``character'' in place of ``person.''
While we do not know the extent to which offenders use these guides as instructional for NCII creation, prior work has found that stated community norms in online spaces can impact user behavior~\cite{fiesler2018reddit, matias2019preventing}. Therefore, the notable absence of clearly established boundaries or acceptable use within the guides suggests that, like other resources in the AIG-SC community~\cite{doringExperiencesAIGeneratedPornography2025, qinThoughtYouWere2026, medeirosCharacterizingResourceSharing2026}, they may be used and shared by offenders.

\header{What Happened to Our Defenses?}
We observe that even though foundation model providers attempt to curb abuse by adding defenses, like input/output filtering~\cite{StabilityAI_TransparencyReport}, these are largely ineffective. For example, we identify an evasion technique, which we term \textit{likeness prompting} (\iref{irisk:prompting-persona}), whereby attackers generate text prompts describing sexualized, identifiable personas such that they could be fed to a model to generate AIG-NCII while evading input/output filters that may only detect the use of a specific individual's name or the upload of content depicting that individual.

Our findings confirm that open-weight models, which are challenging to safeguard once released~\cite{kapoorPositionSocietalImpact2024}, play an essential role (Section \ref{sec:basic-pipeline}) in the technical workflows that enable AIG-NCII, whether that is basic, prompt-only workflows on fine-tuned derivatives of open-weight models or more advanced workflows that adapt base models. 
Further, with increasingly low-resource training techniques (e.g., LoRA), lightweight, hypercustomized models can be trained and used locally to reintroduce intentionally removed concepts (Section~\ref{sec:training}).

While open-weight models are used extensively, we find evidence that both open- and closed-weight models that generate text outputs (i.e., LLMs, VLMs) are also used to lower the barrier to entry for AIG-SC and, subsequently, AIG-NCII, by constructing text inputs that can be fed to models and aiding users trying to set up local content generation workflows (Section \ref{subsec:multi-step}). 
Guides describe tools like ChatGPT as \pquote{the best resource to learn to train} (G172) (Section~\ref{subsec:multi-step}) and describe multi-turn prompting that breaks brittle prompt filters installed in such models. 
We find that, generally, these workflows are becoming easier to use; popular general-purpose generation software (e.g., ComfyUI, A1111) support drag-and-drop workflows that allow more technical users to easily upskill less technical users with ready-to-use workflows. Therefore, although we model an intentional, motivated attacker, this also has serious implications for those who may not intend to produce harm, or are simply curious~\cite{flynn2025sexualized, minkUnlimitedRealmExploration2026}.
Future work could examine to what extent the ease of use and absence of boundaries nudge people to consider AIG-NCII creation who otherwise might not have.

\header{Challenges of Regulating a Dual-Use Ecosystem}
Public scrutiny of NCII creation has primarily focused on nudification apps, which are more clearly intended to support NCII workflows. Accordingly, Minnesota passed a law, put into effect August 1, 2026, that prohibits the use of such nudification apps, described by the bill as any ``website, application, software, program or other service to nudify an image or video, or nudify an image or video on behalf of a user''~\cite{jasminemithaniMinnesotaPassesNations2026, minnesotaNudificationLaw}.
Prior work has characterized platform-level vulnerabilities and provided recommendations for technology owners, such as including clear platform policies and conducting regular red-teaming exercises~\cite{daffalla2026dual, williams2025there, gibson2025analyzing}.
This may be a place to start; in our work, we identify an undressing workflow built with use-case agnostic tools (\iref{irisk:nudification-identity}, \sref{srisk:nudification-sexual}). The inputs (a persona image) and outputs (a nudified image of that same persona) mirror that of a nudification application and therefore could potentially be evaluated in the same way. 
Further, should explicit nudification services be banned, users may turn to guides like those we analyze to replicate the same functionality. 
Our findings underscore the need to study whether bans like that instituted in Minnesota meaningfully reduce abuse at a population level and whether any such effects are obtained without chilling lawful expression~\cite{silverman2026musk}. Either way, the dual-nature of the workflows we document presents a challenge for regulation; these workflows support both abuse and the legitimate generation of sexual content. 
There is no single or set of components that can be outright banned. 
Regulators are left trying to understand how to identify when a workflow ``evolves'' from an AIG-SC workflow to one that enables NCII. This question echoes debates about gun control~\cite{halbrook2024meaning}, specifically asking at what point the individual pieces that compose to create a gun are capable of producing harm. 
As we have demonstrated, the two composable pieces of NCII are sexual concepts and identity concepts. One place to look could be countries like Denmark, that have proposed modifications to copyright law such that individuals have copyrights to their generated ``likeness''~\cite{techPolicyDenmark}. 
Whether and how it will be technically possible to enforce such law, while preserving privacy, is a challenge for which the security and privacy community is well positioned to offer expertise.

\section*{Ethical Considerations}
\label{sec:ethics}
While content analysis of intimate images and video
are established methods of researching gender-based/sexual violence within sexual content~\cite{klaassen2015gender,carrotte2020sexual, bridges2010aggression, barron2000sexual, cowan1994racism, peters2014age}
and is often conducted within academic disciplines such as
psychology~\cite{palys1986testing, cowan1994racism, garcia1991content,sun2008comparison},
media studies~\cite{prince1990power},
gender, sex, and/or porn studies~\cite{klaassen2015gender,bridges2010aggression, barron2000sexual, garcia1991content,fritz2017orgasms, peters2014age},
and health studies~\cite{carrotte2020sexual},
such methods still hold risks.
As such, our research frequently held internal continual discussions for how to address possible risks and their potential impacts towards several relevant stakeholders, including the authors of \civitai{} guides,
those whose identity may (non-consensually) appear within \civitai{} guides,
safety of current and future victim-survivors of sexual violence,
and the members of the research team.
As such, we now detail our ethical considerations and our corresponding procedures for mitigating them.
This study was overseen and deemed exempt by our institution's IRB. All procedures followed the local and federal regulations of the jurisdiction in which each researcher resided.

\header{Privacy and Rights of Potential Victim Survivors, and Those Depicted in Articles}
To handle content responsibly, we gather data manually and do so only by researchers outside a degree-granting program (e.g., not within an undergraduate, master's, or PhD program).
We only capture public guides, intentionally posted by their author on \civitai{}.
While \civitai{}'s policy prohibits all forms of CSAM, and the use of any real identity, generated or otherwise, on their site, we suspect (and find) that prohibited content still remains.
We initially hoped to perform this without human involvement by requesting access to CSAM detection API
but were not able to access commercial tools only available to verified enterprises,
and thus resorted to manual filtering.
As noted, all data was initially gathered by non-student researchers, and initially vetted for prohibited content; only then was it captured and provided to other team members for analysis.
If at any point a member of the team believed they identified CSAM or a real identity, the article would be reported to NCMEC (for CSAM) and CivitAI (for a real identity), and ensured that it was not digitally stored on any device (for CSAM and NCII).\footnote{The team considered systematically reverse-searching all images, but due to potential privacy concerns in such systems~\cite{pratik2023privacy}, decided against their use and instead relied on researcher knowledge.}
In the end, we flagged the following number of posts: 2 to NCMEC for suspected AIG-CSAM, 1 to \civitai{} for encouraging the production of NCII (although it did not contain NCII), and 4 to \civitai{} for containing (non-intimate) depictions of real identities. At the time of writing, three weeks after reporting, only one of the reported guides has been removed from the platform.

\header{Researcher Safety}
Our team consisted of multiple researchers with PhDs, as well as researchers in degree-granting programs.
Every person in a degree-granting program was provided a notice regarding research activities involving materials with intimate content; the notice contains the following information and reminders:
(1) researcher activities may contain intimate content and themes,
(2) researcher participation is voluntary, and they may stop at any time without adverse treatment,
(3) exposure to content could be uncomfortable, anxiety-inducing, or triggering,
(4) the name and contact of support options including psychological services, student conduct, resources for sexual health, Title IX coordinator, and mental health assistant for the project,
(5) expected professional conduct, behavior, and data security,
(6) policies pertaining to, and mechanisms for sexual harassment and retaliation including all PIs, counseling services, Title IX coordinator, and local law enforcement.
Student researchers were checked in on by senior researchers. We also retain a licensed counselor who specializes in trauma \& sexual violence whose services are freely available to anyone on the team.

Beyond collecting data with an anonymous account created solely for the research project, researchers did not interact with any site users.
No other models or datasets were downloaded or run by any researcher.

\header{Publication of Research}
We emphasize that while we report novel academic findings for how NCII is/can be generated with widely distributed tools,
we only report on popular, already taught methods that, as our data show, have been used to generate AIG-NCII.
Furthermore, this work directly enables E.U. safety requirements for AI model and application developers to proactively test for ``reasonably foreseeable misuse''~\cite{euAIAct} by AIG-NCII offenders; we agree and follow a similar rationale. While we agree our work provides an understanding of how abuse is performed, we also believe this is essential for mitigating and reducing ongoing abuse.

\cleardoublepage
\bibliographystyle{plainurl}
\bibliography{references/bibliography, references/zotero}

@inproceedings{debenedetti2024privacy,
  title={Privacy side channels in machine learning systems},
  author={Debenedetti, Edoardo and Severi, Giorgio and Carlini, Nicholas and Choquette-Choo, Christopher A and Jagielski, Matthew and Nasr, Milad and Wallace, Eric and Tram{\`e}r, Florian},
  booktitle={33rd USENIX Security Symposium (USENIX Security 24)},
  pages={6831--6848},
  year={2024}
}

@misc{silverman2026musk,
  author = {Craig Silverman and Alexios Mantzarlis},
  title = {Briefing: Musk Sues Minnesota over Synthetic Nudes Law},
  year = {2026},
  month = jul,
  publisher = {Indicator},
  url = {https://indicator.media/p/briefing-musk-sues-minnesota-over-synthetic-nudes-law}
}

@misc{redmiles2024nist,
  author = {Elissa M. Redmiles and Sarah Adel Bargal and Natalie Grace Brigham and Nina Grgi{\'c}-Hla{\v{c}}a and Tadayoshi Kohno and Jaron Mink and Veronica A. Rivera and Carmela Troncoso and Lucy Qin and Miranda Wei},
  title = {Georgetown Massive Data Institute and Department of Computer Science Response to Federal Register Document 2023-28232, {``Request for Information (RFI) Related to NIST's Assignments Under Sections 4.1, 4.5 and 11 of the Executive Order Concerning Artificial Intelligence''}},
  year = {2024},
  month = feb,
  publisher = {Massive Data Institute and Department of Computer Science, Georgetown University},
  address = {Washington, DC},
  note = {Comment no. 9},
  url = {https://mdi.georgetown.edu/news/formal-response-mdi-comment-on-nist-ai-executive-order-feb2024/}
}

@inproceedings{ruge2020frankenstein,
  title={Frankenstein: Advanced wireless fuzzing to exploit new bluetooth escalation targets},
  author={Ruge, Jan and Classen, Jiska and Gringoli, Francesco and Hollick, Matthias},
  booktitle={29th USENIX Security Symposium (USENIX Security 20)},
  pages={19--36},
  year={2020}
}

@inproceedings{bernhard2024darthshader,
  title={DarthShader: Fuzzing WebGPU shader translators \& compilers},
  author={Bernhard, Lukas and Schiller, Nico and Schloegel, Moritz and Bars, Nils and Holz, Thorsten},
  booktitle={Proceedings of the 2024 on ACM SIGSAC Conference on Computer and Communications Security},
  pages={690--704},
  year={2024}
}

@misc{publicity_law,
  title        = "{publicity | Wex | US Law | LII / Legal Information Institute}",
  note = {\url{https://www.law.cornell.edu/wex/publicity}},
  organization = "Legal Information Institute, Cornell Law School"
}

@misc{maiberg25_civitai_people_ban,
  author       = "Emanuel Maiberg",
  title        = "{Civitai Ban of Real People Content Deals Major Blow to the Nonconsensual AI Porn Ecosystem}",
  note         = {\url{https://www.404media.co/civitai-ban-of-real-people-content-deals-major-blow-to-the-nonconsensual-ai-porn-ecosystem/}},
  year         = "2025",
  howpublished = "404 Media (online)",
  month        = may
}

@misc{maiberg25_civitai_cutoff,
  author       = "Emanuel Maiberg",
  title        = "{Civitai, Site Used to Generate AI Porn, Cut Off by Credit Card Processor}",
  note         = {\url{https://www.404media.co/civitai-site-used-to-generate-ai-porn-cut-off-by-credit-card-processor/}},
  year         = "2025",
  howpublished = "404 Media (online)",
  month        = may
}

@misc{techPolicyDenmark,
  author={Giovana Fleck},
  title="{Denmark Leads EU Push to Copyright Faces in Fight Against Deepfakes
}",
  note={\url{https://www.techpolicy.press/denmark-leads-eu-push-to-copyright-faces-in-fight-against-deepfakes/}},
  year="2025",
  organization = "Tech Policy Press"
}

@article{azungah2018qualitative,
  title={{Qualitative Research: Deductive and Inductive Approaches to Data Analysis}},
  author={Azungah, Theophilus},
  journal={Qualitative Research Journal},
  volume={18},
  number={4},
  pages={383--400},
  year={2018},
  publisher={Emerald Publishing Limited}
}

@article{schreyer2015overview,
  title={An overview of legal protection for fictional characters: Balancing public and private interests},
  author={Schreyer, Amanda},
  journal={Cybaris Intell. Prop. L. Rev.},
  volume={6},
  pages={50},
  year={2015},
  publisher={HeinOnline}
}

@misc{maiberg24_ai_pimping,
  author       = "Jason Koebler",
  title        = "{‘AI Instagram Influencers’ Are Deepfaking Their Faces Onto Real Women’s Bodies}",
  note         = {\url{https://www.404media.co/ai-influencers-are-deepfaking-their-faces-onto-real-womens-bodies/}},
  year         = "2024",
  howpublished = "404 Media (online)",
  month        = apr
}

@misc{conversation2015_reddit_revenge_porn,
  author       = "Emily van der Nagel and James Meese",
  title        = "{Reddit tackles ‘revenge porn’ and celebrity nudes}",
  note         = {\url{https://theconversation.com/reddit-tackles-revenge-porn-and-celebrity-nudes-38112}},
  year         = "2015",
  howpublished = "The Conversation (online)",
  month        = feb
}

@article{flynn2025sexualized,
  title={Sexualized deepfake abuse: Perpetrator and victim perspectives on the motivations and forms of non-consensually created and shared sexualized deepfake imagery},
  author={Flynn, Asher and Powell, Anastasia and Eaton, Asia A and Scott, Adrian J},
  journal={Journal of Interpersonal Violence},
  year={2025},
  publisher={SAGE Publications}
}

@misc{StabilityAI_TransparencyReport,
  author       = {{Stability AI}},
  title        = {{Stability AI’s Annual Integrity Transparency Report
}},
  year         = {2025},
  month        = {April},
  note         = {\url{https://stability.ai/news-updates/stability-ais-annual-integrity-transparency-report}}
}

@misc{StableSafety,
  author       = {Stability AI},
  title        = {{Stable Safety — Stability AI}},
  year         = {},
  month        = {},
  note         = {\url{https://stability.ai/safety}}
}

@article{carrotte2020sexual,
  title={Sexual behaviors and violence in pornography: Systematic review and narrative synthesis of video content analyses},
  author={Carrotte, Elise R and Davis, Angela C and Lim, Megan SC},
  journal={Journal of medical internet research},
  volume={22},
  number={5},
  pages={e16702},
  year={2020},
  publisher={JMIR Publications Toronto, Canada}
}

@article{klaassen2015gender,
  title={Gender (in) equality in internet pornography: A content analysis of popular pornographic internet videos},
  author={Klaassen, Marleen JE and Peter, Jochen},
  journal={The Journal of Sex Research},
  volume={52},
  number={7},
  pages={721--735},
  year={2015},
  publisher={Taylor \& Francis}
}

@article{sun2008comparison,
  title={A comparison of male and female directors in popular pornography: What happens when women are at the helm?},
  author={Sun, Chyng and Bridges, Ana and Wosnitzer, Robert and Scharrer, Erica and Liberman, Rachael},
  journal={Psychology of Women Quarterly},
  volume={32},
  number={3},
  pages={312--325},
  year={2008},
  publisher={SAGE Publications Sage CA: Los Angeles, CA}
}

@article{peters2014age,
  title={Age is in the eye of the beholder: Examining the cues employed to construct the illusion of youth in teen pornography},
  author={Peters, Evyn M and Morrison, Todd and McDermott, Daragh T and Bishop, CJ and Kiss, Mark},
  journal={Sexuality \& Culture},
  volume={18},
  number={3},
  pages={527--546},
  year={2014},
  publisher={Springer}
}

@article{fritz2017orgasms,
  title={From orgasms to spanking: A content analysis of the agentic and objectifying sexual scripts in feminist, for women, and mainstream pornography},
  author={Fritz, Niki and Paul, Bryant},
  journal={Sex Roles},
  volume={77},
  number={9},
  pages={639--652},
  year={2017},
  publisher={Springer}
}

@article{prince1990power,
  title={Power and pain: Content analysis and the ideology of pornography},
  author={Prince, Stephen},
  journal={Journal of Film and Video},
  volume={42},
  number={2},
  pages={31--41},
  year={1990},
  publisher={JSTOR}
}

@article{garcia1991content,
  title={A content analysis of erotic videos},
  author={Garcia, Luis T and Milano, Laureen},
  journal={Journal of Psychology \& Human Sexuality},
  volume={3},
  number={2},
  pages={95--103},
  year={1991},
  publisher={Taylor \& Francis}
}

@article{cowan1994racism,
  title={Racism and sexism in interracial pornography a content analysis},
  author={Cowan, Gloria and Campbell, Robin R},
  journal={Psychology of Women Quarterly},
  volume={18},
  number={3},
  pages={323--338},
  year={1994},
  publisher={Wiley Online Library}
}

@article{palys1986testing,
  title={Testing the common wisdom: The social content of video pornography.},
  author={Palys, Ted S},
  journal={Canadian Psychology/Psychologie Canadienne},
  volume={27},
  number={1},
  pages={22},
  year={1986},
  publisher={Canadian Psychological Association}
}

@article{barron2000sexual,
  title={Sexual violence in three pornographic media: Toward a sociological explanation},
  author={Barron, Martin and Kimmel, Michael},
  journal={Journal of Sex Research},
  volume={37},
  number={2},
  pages={161--168},
  year={2000},
  publisher={Taylor \& Francis}
}

@article{bridges2010aggression,
  title={Aggression and sexual behavior in best-selling pornography videos: A content analysis update},
  author={Bridges, Ana J and Wosnitzer, Robert and Scharrer, Erica and Sun, Chyng and Liberman, Rachael},
  journal={Violence against women},
  volume={16},
  number={10},
  pages={1065--1085},
  year={2010},
  publisher={Sage Publications Sage CA: Los Angeles, CA}
}

@misc{civitaiControlNet,
  title={Civitai Guide to ControlNet – Part I – Basics},
  author={CivitAI},
  url={https://education.civitai.com/civitai-guide-to-controlnet/},
  note={Accessed: 2026-08-25}
}

@misc{civitaiSafety,
  title={Safety Center},
  url={https://civitai.com/safety},
  note={Accessed: 2026-05-19}
}

@misc{code6851,
  title={15 U.S. Code § 6851 - Civil action relating to disclosure of intimate images},
  url={https://www.law.cornell.edu/uscode/text/15/6851},
  note={Accessed: 2026-06-15}
}

@misc{takeItDown,
  title = {ICYMI: President Trump Signs TAKE IT DOWN Act into Law},
  url={https://www.whitehouse.gov/releases/2025/05/icymi-president-trump-signs-take-it-down-act-into-law/},
  author = {The White House},
  month = {},
  year = {2025},
  note = {Accessed: 2026-05-05}
}

@misc{minnesotaNudificationLaw,
  title = {New law bans access to nudification technology},
  url={https://www.house.mn.gov/NewLaws/story/2026/5741},
  author = {Minnesota House of Representatives},
  month = {},
  year = {2026},
  note = {Accessed: 2026-08-07}
}

@misc{EUai,
  title = {EU AI Act: first regulation on artificial intelligence
},
  url={https://www.europarl.europa.eu/topics/en/article/20230601STO93804/eu-ai-act-first-regulation-on-artificial-intelligence},
  author = {European Parliament},
  month = {},
  year = {},
  note = {Accessed: 2026-05-05}
}

@misc{civitaiPolicy,
  title = {Policy Update: Removal of Real-Person Likeness Content | Civitai},
  url={https://civitai.com/articles/15022/policy-update-removal-of-real-person-likeness-content},
  author = {CivitaiOfficial},
  month = {May},
  year = {2025},
  note = {Accessed: 2026-05-05}
}

@inproceedings{ghosh2026marketplace,
  title={A Marketplace for AI-Generated Adult Content and Deepfakes},
  author={Ghosh, Shalmoli and DeVerna, Matthew R and Menczer, Filippo},
  booktitle={The 2026 ACM Conference on Fairness, Accountability, and Transparency},
  pages={3466--3485},
  year={2026}
}

@article{hu2021lora,
  title={Lora: Low-rank adaptation of large language models},
  author={Hu, Edward J and Shen, Yelong and Wallis, Phillip and Allen-Zhu, Zeyuan and Li, Yuanzhi and Wang, Shean and Wang, Lu and Chen, Weizhu},
  journal={arXiv preprint arXiv:2106.09685},
  year={2021}
}

@inproceedings{gibson2025analyzing,
  title={Analyzing the AI Nudification Application Ecosystem},
  author={Gibson, Cassidy and Olszewski, Daniel and Brigham, Natalie Grace and Crowder, Anna and Butler, Kevin RB and Traynor, Patrick and Redmiles, Elissa M and Kohno, Tadayoshi},
  booktitle={34th USENIX Security Symposium (USENIX Security 25)},
  pages={1--20},
  year={2025}
}

@article{williams2025there,
  title={“There Are No Limits!”: AI Undressing Apps and the Normalization of Nonconsensual Intimate Deepfakes},
  author={Williams, Kaylee},
  journal={Violence Against Women},
  pages={10778012251397966},
  year={2025},
  publisher={SAGE Publications Sage CA: Los Angeles, CA}
}

@misc{ponyRef,
  title={Pony Diffusion V6 XL},
  author = {},
  month = {},
  year = {2026},
  note = {Accessed: 2026-05-28}
}

@misc{noobRef,
  title={NoobAI},
  url={https://civitai.com/ecosystems/noobai},
  author = {Laxhar Lab},
  month = {},
  year = {},
  note = {Accessed: 2026-08-23}
}

@inproceedings{goller2026systematic,
  title={Systematic Analysis of the Unintentional CSAM-Generation-Potential of Text-to-Image Models},
  author={G{\"o}ller, Nicolas and Steinebach, Martin},
  booktitle={Proceedings of the IEEE/CVF Winter Conference on Applications of Computer Vision},
  pages={493--502},
  year={2026}
}

@article{wan2025wan,
  title={Wan: Open and advanced large-scale video generative models},
  author={Wan, Team and Wang, Ang and Ai, Baole and Wen, Bin and Mao, Chaojie and Xie, Chen-Wei and Chen, Di and Yu, Feiwu and Zhao, Haiming and Yang, Jianxiao and others},
  journal={arXiv preprint arXiv:2503.20314},
  year={2025}
}

@inproceedings{zhang2023adding,
  title={Adding conditional control to text-to-image diffusion models},
  author={Zhang, Lvmin and Rao, Anyi and Agrawala, Maneesh},
  booktitle={Proceedings of the IEEE/CVF international conference on computer vision},
  pages={3836--3847},
  year={2023}
}

@article{ye2023ip,
  title={IP-Adapter: Text compatible image prompt adapter for text-to-image diffusion models},
  author={Ye, Hu and Zhang, Jun and Liu, Sibo and Han, Xiao and Yang, Wei},
  journal={arXiv preprint arXiv:2308.06721},
  year={2023}
}

@inproceedings{podell2024sdxl,
  title={Sdxl: Improving latent diffusion models for high-resolution image synthesis},
  author={Podell, Dustin and English, Zion and Lacey, Kyle and Blattmann, Andreas and Dockhorn, Tim and M{\"u}ller, Jonas and Penna, Joe and Rombach, Robin},
  booktitle={International Conference on Learning Representations},
  volume={2024},
  pages={1862--1874},
  year={2024}
}

@misc{danbooruRef,
  title={Danbooru},
  url = {https://danbooru.donmai.us/},
  note = {This links to a website that contains sexual content. Accessed: 06-09-2026}
}

@article{mcdonald2019reliability,
  title={Reliability and inter-rater reliability in qualitative research: Norms and guidelines for CSCW and HCI practice},
  author={McDonald, Nora and Schoenebeck, Sarita and Forte, Andrea},
  journal={Proceedings of the ACM on human-computer interaction},
  volume={3},
  number={CSCW},
  pages={1--23},
  year={2019},
  publisher={ACM New York, NY, USA}
}

@article{halbrook2024meaning,
  title={The Meaning of ``{Firearm}'' and ``{Frame or Receiver}'' in the {Federal Gun Control Act}: {ATF}'s 2022 {Final Rule} in Light of Text, Precedent, and History},
  author={Halbrook, Stephen P},
  month        = {September},
  year         = {2024},
  url = {https://papers.ssrn.com/sol3/papers.cfm?abstract_id=4928892}
}

@inproceedings{pratik2023privacy,
  title={Privacy protection against reverse image search},
  author={Pratik, Rishank and Sendhil, R},
  booktitle={2023 Third International Conference on Artificial Intelligence and Smart Energy (ICAIS)},
  pages={1207--1214},
  year={2023},
  organization={IEEE}
}

@article{daffalla2026dual,
  title={Dual-Use AI Face Swap Apps Are Mostly Unsafe: A Systematic Safety Audit},
  author={Daffalla, Alaa and Chao, Sarah and Zeng, Eric},
  journal={arXiv preprint arXiv:2605.24735},
  year={2026}
}

@misc{euAIAct,
  author       = {{Future of Life Institute}},
  title        = {Article 53: Obligations for Providers of General-Purpose {AI} Models},
  year         = {2024},
  organization = {EU Artificial Intelligence Act},
  url          = {https://artificialintelligenceact.eu/article/53/}
}

@inproceedings{han2025characterizing,
  title={Characterizing the MrDeepFakes Sexual Deepfake Marketplace},
  author={Han, Catherine and Li, Anne and Kumar, Deepak and Durumeric, Zakir},
  booktitle={34th USENIX Security Symposium (USENIX Security 25)},
  pages={5169--5188},
  year={2025}
}

@article{cui2026celebrities,
  title={From Celebrities to Anyone: Characterizing AI Nudification Content, Technology, and Community Dynamics on 4chan},
  author={Cui, Chi and Wu, Yixin and Zhang, Yang},
  journal={arXiv preprint arXiv:2606.27234},
  year={2026}
}

@article{cuevas2026deepfake,
  title={Deepfake Pornography is Resilient to Regulatory and Platform Shocks},
  author={Cuevas, Alejandro and Ribeiro, Manoel Horta},
  journal={arXiv preprint arXiv:2602.02754},
  year={2026}
}

@article{jasminemithaniAdvocates,
  address = {Washington, DC},
  title = {`Nudify' apps are still available at the click of a button},
  url = {https://19thnews.org/2026/05/letter-state-attorneys-general-nudify-apps-accountable/},
  journal = {The 19th News},
  author = {Jasmine Mithani},
  month = may,
  year = {2026}
}

@inproceedings{fiesler2018reddit,
  title={Reddit rules! characterizing an ecosystem of governance},
  author={Fiesler, Casey and Jiang, Jialun and McCann, Joshua and Frye, Kyle and Brubaker, Jed},
  booktitle={Proceedings of the International AAAI Conference on Web and Social Media},
  volume={12},
  number={1},
  year={2018}
}

@article{matias2019preventing,
  title={Preventing harassment and increasing group participation through social norms in 2,190 online science discussions},
  author={Matias, J Nathan},
  journal={Proceedings of the National Academy of Sciences},
  volume={116},
  number={20},
  pages={9785--9789},
  year={2019},
  publisher={National Academy of Sciences}
}

@incollection{martin2022image,
  title={Image-based sexual abuse and deepfakes: A survivor turned activist’s perspective},
  author={Martin, Noelle},
  booktitle={The Palgrave handbook of gendered violence and technology},
  pages={55--72},
  year={2022},
  publisher={Springer}
}

@article{yang2025many,
  title={Many-turn jailbreaking},
  author={Yang, Xianjun and Xiao, Liqiang and Li, Shiyang and Ladhak, Faisal and Yun, Hyokun and Petzold, Linda Ruth and Xu, Yi and Wang, William Yang},
  journal={arXiv preprint arXiv:2508.06755},
  year={2025}
}

@article{song2026multibreak,
  title={MultiBreak: A Scalable and Diverse Multi-turn Jailbreak Benchmark for Evaluating LLM Safety},
  author={Song, Jialin and Liu, Xiaodong and Yang, Weiwei and Chen, Wuyang and Feng, Mingqian and Zhu, Xuekai and Gao, Jianfeng},
  booktitle = {{ICML}},
  year={2026}
}

@article{schuhmann2022laion,
  title={LAION-5B: An open large-scale dataset for training next generation image-text models},
  author={Schuhmann, Christoph and Beaumont, Romain and Vencu, Richard and Gordon, Cade and Wightman, Ross and Cherti, Mehdi and Coombes, Theo and Katta, Aarush and Mullis, Clayton and Wortsman, Mitchell and others},
  journal={Advances in neural information processing systems},
  volume={35},
  pages={25278--25294},
  year={2022}
}

@techreport{europeansexworkersrightsallianceedeepfakeswork26,
  title = {Deepfakes and Sex Work: Harms, Benefits, and Worker-Led Recommendations},
  url = {https://www.eswalliance.org/deepfakes_2026},
  author = {{European Sex Workers' Rights Alliance (ESWA)}},
  month = jul,
  year = {2026}
}

@inproceedings{redmon2016you,
  title={You only look once: Unified, real-time object detection},
  author={Redmon, Joseph and Divvala, Santosh and Girshick, Ross and Farhadi, Ali},
  booktitle={Proceedings of the IEEE conference on computer vision and pattern recognition},
  pages={779--788},
  year={2016}
}

@inproceedings{liu2022compositional,
  title={Compositional visual generation with composable diffusion models},
  author={Liu, Nan and Li, Shuang and Du, Yilun and Torralba, Antonio and Tenenbaum, Joshua B},
  booktitle={European conference on computer vision},
  pages={423--439},
  year={2022},
  organization={Springer}
}

@article{yuan2023inserting,
  title={Inserting Anybody in Diffusion Models via Celeb Basis},
  author={Yuan, Ge and Cun, Xiaodong and Zhang, Yong and Li, Maomao and Qi, Chenyang and Wang, Xintao and Shan, Ying and Zheng, Huicheng},
  year={2023},
  booktitle = {{NeurIPS}},
}

@inproceedings{li2024safegen,
  title={Safegen: Mitigating sexually explicit content generation in text-to-image models},
  author={Li, Xinfeng and Yang, Yuchen and Deng, Jiangyi and Yan, Chen and Chen, Yanjiao and Ji, Xiaoyu and Xu, Wenyuan},
  booktitle={Proceedings of the 2024 on ACM SIGSAC Conference on Computer and Communications Security},
  pages={4807--4821},
  year={2024}
}

@misc{debenedetti2023privacy,
  author = {Edoardo Debenedetti and Florian Tram{\`e}r},
  title = {Privacy Side Channels in Machine Learning Systems},
  year = {2023},
  month = sep,
  howpublished = {\url{https://spylab.ai/blog/side-channels-machine-learning/}},
  note = {SPY Lab}
}

@article{vitisPrivateHiddenObscured2020,
  title = {Private, {Hidden} and {Obscured}: {Image}-{Based} {Sexual} {Abuse} in {Singapore}},
  volume = {15},
  url = {http://link.springer.com/10.1007/s11417-019-09293-0},
  doi = {10.1007/s11417-019-09293-0},
  number = {1},
  journal = {Asian Journal of Criminology},
  author = {Vitis, Laura},
  month = mar,
  year = {2020},
  pages = {25--43}
}

@article{mcglynnRevengePornContinuum2017,
  title = {Beyond ‘{Revenge} {Porn}’: {The} {Continuum} of {Image}-{Based} {Sexual} {Abuse}},
  volume = {25},
  url = {http://link.springer.com/10.1007/s10691-017-9343-2},
  doi = {10.1007/s10691-017-9343-2},
  number = {1},
  journal = {Feminist Legal Studies},
  author = {McGlynn, Clare and Rackley, Erika and Houghton, Ruth},
  month = apr,
  year = {2017},
  pages = {25--46}
}

@article{huberShadowMeOld2023,
  title = {‘{A} shadow of me old self’: {The} impact of image-based sexual abuse in a digital society},
  volume = {29},
  url = {https://doi.org/10.1177/02697580211063659},
  doi = {10.1177/02697580211063659},
  number = {2},
  journal = {International Review of Victimology},
  author = {Huber, Antoinette},
  year = {2023},
  note = {\_eprint: https://doi.org/10.1177/02697580211063659},
  pages = {199--216}
}

@article{batesRevengePornMental2017,
  title = {Revenge {Porn} and {Mental} {Health}: {A} {Qualitative} {Analysis} of the {Mental} {Health} {Effects} of {Revenge} {Porn} on {Female} {Survivors}},
  volume = {12},
  number = {1},
  journal = {Feminist Criminology},
  author = {Bates, Samantha},
  month = jan,
  year = {2017},
  pages = {22--42}
}

@inproceedings{freedStalkersParadiseHow2018,
  address = {Montreal QC Canada},
  title = {“{A} {Stalker}'s {Paradise}”: {How} {Intimate} {Partner} {Abusers} {Exploit} {Technology}},
  url = {https://dl.acm.org/doi/10.1145/3173574.3174241},
  doi = {10.1145/3173574.3174241},
  booktitle = {Proceedings of the 2018 {CHI} {Conference} on {Human} {Factors} in {Computing} {Systems}},
  publisher = {ACM},
  author = {Freed, Diana and Palmer, Jackeline and Minchala, Diana and Levy, Karen and Ristenpart, Thomas and Dell, Nicola},
  month = apr,
  year = {2018},
  pages = {1--13}
}

@article{emanuelmaibergInstagramsNudifyAds2024,
  title = {Instagram's {Nudify} {Ads}},
  url = {https://www.404media.co/email/d2bebba9-5808-44fc-8352-d93d1791a5ff/?ref=daily-stories-newsletter},
  journal = {404 Media},
  author = {Emanuel Maiberg},
  month = apr,
  year = {2024}
}

@article{emanuelmaibergAppleRemovesNonconsensual2024,
  title = {Apple {Removes} {Nonconsensual} {AI} {Nude} {Apps} {Following} 404 {Media} {Investigation}},
  url = {https://www.404media.co/apple-removes-nonconsensual-ai-nude-apps-following-404-media-investigation/},
  journal = {404 Media},
  author = {{Emanuel Maiberg}},
  month = apr,
  year = {2024}
}

@article{huberNonconsensualIntimateImage2024,
  title = {Non-consensual intimate image distribution: {Nature}, removal, and implications for the {Online} {Safety} {Act}},
  url = {https://journals.sagepub.com/doi/10.1177/14773708241255821},
  doi = {10.1177/14773708241255821},
  journal = {European Journal of Criminology},
  author = {Huber, Antoinette Raffaela and Ward, Zara},
  month = jul,
  year = {2024},
  pages = {14773708241255821}
}

@inproceedings{qinDidTheyFing2024,
  address = {Philadelphia, PA},
  title = {"{Did} {They} {F}***ing {Consent} to {That}?": {Safer} {Digital} {Intimacy} via {Proactive} {Protection} {Against} {Image}-{Based} {Sexual} {Abuse}},
  url = {https://www.usenix.org/conference/usenixsecurity24/presentation/qin},
  booktitle = {33rd {USENIX} {Security} {Symposium} ({USENIX} {Security} 24)},
  publisher = {USENIX Association},
  author = {Qin, Lucy and Hamilton, Vaughn and Wang, Sharon and Aydinalp, Yigit and Scarlett, Marin and Redmiles, Elissa M.},
  month = aug,
  year = {2024},
  pages = {55--72}
}

@inproceedings{batoolExpandingConceptsNonConsensual2024,
  address = {Honolulu HI USA},
  title = {Expanding {Concepts} of {Non}-{Consensual} {Image}-{Disclosure} {Abuse}: {A} {Study} of {NCIDA} in {Pakistan}},
  url = {https://dl.acm.org/doi/10.1145/3613904.3642871},
  doi = {10.1145/3613904.3642871},
  booktitle = {Proceedings of the {CHI} {Conference} on {Human} {Factors} in {Computing} {Systems}},
  publisher = {ACM},
  author = {Batool, Amna and Naseem, Mustafa and Toyama, Kentaro},
  month = may,
  year = {2024},
  pages = {1--17}
}

@article{henryImageBasedSexualAbuse2023,
  title = {Image-{Based} {Sexual} {Abuse} as a {Means} of {Coercive} {Control}: {Victim}-{Survivor} {Experiences}},
  volume = {29},
  url = {https://journals.sagepub.com/doi/10.1177/10778012221114918},
  doi = {10.1177/10778012221114918},
  number = {6-7},
  journal = {Violence Against Women},
  author = {Henry, Nicola and Gavey, Nicola and Johnson, Kelly},
  month = may,
  year = {2023},
  pages = {1206--1226}
}

@article{gehlTrainingComputersSee2017,
  title = {Training {Computers} to {See} {Internet} {Pornography}: {Gender} and {Sexual} {Discrimination} in {Computer} {Vision} {Science}},
  volume = {18},
  url = {http://journals.sagepub.com/doi/10.1177/1527476416680453},
  doi = {10.1177/1527476416680453},
  number = {6},
  journal = {Television \& New Media},
  author = {Gehl, Robert W. and Moyer-Horner, Lucas and Yeo, Sara K.},
  month = sep,
  year = {2017},
  pages = {529--547}
}

@inproceedings{hawkinsDeepfakesDemandRise2025,
  address = {New York, NY, USA},
  series = {{FAccT} '25},
  title = {Deepfakes on {Demand}: {The} rise of accessible non-consensual deepfake image generators},
  url = {https://dl.acm.org/doi/10.1145/3715275.3732107},
  doi = {10.1145/3715275.3732107},
  booktitle = {Proceedings of the 2025 {ACM} {Conference} on {Fairness}, {Accountability}, and {Transparency}},
  publisher = {Association for Computing Machinery},
  author = {Hawkins, Will and Mittelstadt, Brent and Russell, Chris},
  month = jun,
  year = {2025},
  pages = {1602--1614}
}

@inproceedings{weiExploringUseAbusive2024,
  title = {Exploring the use of abusive generative {AI} models on {Civitai}},
  booktitle = {Proceedings of the 32nd {ACM} {International} {Conference} on {Multimedia}},
  author = {Wei, Yiluo and Zhu, Yiming and Hui, Pan and Tyson, Gareth},
  year = {2024},
  pages = {6949--6958}
}

@article{henryDevastatingItBroke2023,
  title = {‘{Devastating}, like it broke me’: {Responding} to image-based sexual abuse in {Aotearoa} {New} {Zealand}},
  volume = {23},
  url = {http://journals.sagepub.com/doi/10.1177/17488958221097276},
  doi = {10.1177/17488958221097276},
  number = {5},
  journal = {Criminology \& Criminal Justice},
  author = {Henry, Nicola and Gavey, Nicola and McGlynn, Clare and Rackley, Erika},
  month = nov,
  year = {2023},
  pages = {861--879}
}

@inproceedings{riccioExposedErasedAlgorithmic2024a,
  title = {Exposed or erased: algorithmic censorship of nudity in art},
  booktitle = {Proceedings of the 2024 {CHI} {Conference} on {Human} {Factors} in {Computing} {Systems}},
  author = {Riccio, Piera and Hofmann, Thomas and Oliver, Nuria},
  year = {2024},
  pages = {1--17}
}

@article{emanuelmaibergAIPornMarketplace2023,
  title = {Inside the {AI} {Porn} {Marketplace} {Where} {Everything} and {Everyone} {Is} for {Sale}},
  url = {https://www.404media.co/inside-the-ai-porn-marketplace-where-everything-and-everyone-is-for-sale/},
  journal = {404 Media},
  author = {{Emanuel Maiberg}},
  month = aug,
  year = {2023}
}

@misc{tedcruzS146TAKEIT2025,
  title = {S.146 - {TAKE} {IT} {DOWN} {Act}},
  url = {https://www.congress.gov/bill/119th-congress/senate-bill/146},
  author = {Cruz, Ted},
  month = may,
  year = {2025}
}

@inproceedings{geengUsableSexurityStudying2020,
  title = {Usable Sexurity: Studying People’s Concerns and Strategies When Sexting},
  booktitle = {Sixteenth {Symposium} on {Usable} {Privacy} and {Security} ({SOUPS} 2020)},
  author = {Geeng, Christine and Hutson, Jevan and Roesner, Franziska},
  year = {2020},
  pages = {127--144}
}

@article{butlerSexWorkerHuman2025,
  title = {Sex worker human digital twins and intellectual property: on collaborative futures for sex thinkers, technologists, and lawyers},
  journal = {Porn Studies},
  publisher = {Taylor \& Francis},
  author = {Butler, Yvette and Egwuatu, Chibundo and Canham, Matthew and Sawyer, Ben D},
  year = {2025},
  pages = {1--16}
}

@article{wiseMajorDeepfakePorn2025,
  title = {Major deepfake porn site shuts down},
  url = {https://www.npr.org/2025/05/06/nx-s1-5388422/mr-deepfakes-porn-site-ai-shut-down},
  journal = {NPR News},
  author = {Wise, Alana},
  month = may,
  year = {2025}
}

@article{braunUsingThematicAnalysis2006,
  title = {Using thematic analysis in psychology},
  volume = {3},
  number = {2},
  journal = {Qualitative research in psychology},
  publisher = {Taylor \& Francis},
  author = {Braun, Virginia and Clarke, Victoria},
  year = {2006},
  pages = {77--101}
}

@misc{nationalcenterformissing&exploitedchildrenNationalCenterMissing2026,
  type = {Website},
  title = {National {Center} for {Missing} \& {Exploited} {Children}},
  url = {https://www.missingkids.org/home?lang=en-US},
  author = {{National Center for Missing \& Exploited Children}},
  year = {2026}
}

@inproceedings{princessacintaqiaStopNonconsensualUse2025,
  title = {Stop the {Nonconsensual} {Use} of {Nude} {Images} in {Research}},
  url = {https://openreview.net/forum?id=Ev5xwr3vWh},
  booktitle = {The {Thirty}-{Ninth} {Annual} {Conference} on {Neural} {Information} {Processing} {Systems}},
  author = {Cintaqia, Princessa and Arya, Arshia and Redmiles, Elissa M. and Kumar, Deepak and McDonald, Allison and Qin, Lucy},
  year = {2025}
}

@misc{minkUnlimitedRealmExploration2026,
  title = {"{Unlimited} {Realm} of {Exploration} and {Experimentation}": {Methods} and {Motivations} of {AI}-{Generated} {Sexual} {Content} {Creators}},
  url = {https://dl.acm.org/doi/10.1145/3805689.3812314},
  doi = {10.1145/3805689.3812314},
  author = {Mink, Jaron and Qin, Lucy and Redmiles, Elissa M.},
  note = {Version Number: 1},
  series = {{FAccT} '26},
  booktitle = {Proceedings of the 2026 {ACM} {Conference} on {Fairness}, {Accountability}, and {Transparency}},
  publisher = {Association for Computing Machinery},
  month = jun,
  year = {2026},
  pages = {3242--3273}
}

@article{doringExperiencesAIGeneratedPornography2025,
  title = {Experiences with {AI}-{Generated} {Pornography}: {A} {Quantitative} {Content} {Analysis} of {Reddit} {Posts}},
  url = {https://link.springer.com/10.1007/s10508-025-03227-x},
  doi = {10.1007/s10508-025-03227-x},
  journal = {Archives of Sexual Behavior},
  author = {Döring, Nicola and Le, Thuy Dung and Miller, Dan J.},
  month = sep,
  year = {2025}
}

@article{lapointeGovernanceAIgeneratedPornography2026,
  title = {The governance of {AI}-generated pornography platforms: {A} content analysis},
  url = {https://journals.sagepub.com/doi/10.1177/14614448261421873},
  doi = {10.1177/14614448261421873},
  journal = {New Media \& Society},
  author = {Lapointe, Valerie A and Dubé, Simon and Petit, Aurélie and Kessai, Tinhinane and Rukhlyadyev, Sophia and Gravel, Vivianne and Lafortune, David},
  month = feb,
  year = {2026},
  pages = {14614448261421873}
}

@article{lapointePresentFutureAdult2025,
  title = {The {Present} and {Future} of {Adult} {Entertainment}: {A} {Content} {Analysis} of {AI}-{Generated} {Pornography} {Websites}},
  url = {https://link.springer.com/10.1007/s10508-025-03099-1},
  doi = {10.1007/s10508-025-03099-1},
  journal = {Archives of Sexual Behavior},
  author = {Lapointe, Valerie A. and Dubé, Simon and Rukhlyadyev, Sophia and Kessai, Tinhinane and Lafortune, David},
  month = mar,
  year = {2025}
}

@inproceedings{dawoudUndergroundMainstreamMarketplaces2026,
  title = {From {Underground} to {Mainstream} {Marketplaces}: {Measuring} {AI}-{Enabled} {NSFW} {Deepfakes} on {Fiverr}},
  booktitle = {Symposium on {Usable} {Security} and {Privacy} ({USEC})},
  author = {Dawoud, Mohamed Moustafa and Cuevas, Alejandro and Raman, Ram Sundara},
  year = {2026}
}

@inproceedings{brighamExaminingRisksAI2026,
  author = {Brigham, Natalie Grace and Qin, Lucy and Kohno, Tadayoshi},
  title     = {Examining Risks Through a Characterization of the {AI} Companion Application Ecosystem: A Stratified Sample from the {Apple App Store} and {Google Play Store}},
  booktitle = {Proceedings of the AAAI/ACM Conference on AI, Ethics, and Society},
  volume    = {9},
  number    = {1},
  pages     = {498--511},
  year      = {2026},
  month     = oct,
  doi       = {10.1609/aies.v9i1.43158}
}

@misc{medeirosCharacterizingResourceSharing2026,
  title = {Characterizing {Resource} {Sharing} {Practices} on {Underground} {Internet} {Forum} {Synthetic} {Non}-{Consensual} {Intimate} {Image} {Content} {Creation} {Communities}},
  url = {https://arxiv.org/abs/2604.12190},
  doi = {10.48550/ARXIV.2604.12190},
  publisher = {arXiv},
  author = {Medeiros, Bernardo B. P. and Jadhav, Malvika and Lu, Allison and Kohno, Tadayoshi and Bindschaedler, Vincent and Butler, Kevin R. B.},
  year = {2026},
  note = {Version Number: 1}
}

@misc{wagnerPerpetuatingMisogynyGenerative2025,
  title = {Perpetuating {Misogyny} with {Generative} {AI}: {How} {Model} {Personalization} {Normalizes} {Gendered} {Harm}},
  url = {https://journals.sagepub.com/doi/10.1177/20539517261467361},
  doi = {10.1177/20539517261467361},
  author = {Wagner, Laura and Cetinic, Eva},
  month = may,
  journal={Big Data & Society},
  year={2026},
  publisher={SAGE Publications}
}

@inproceedings{kapoorPositionSocietalImpact2024,
  series = {Proceedings of {Machine} {Learning} {Research}},
  title = {Position: {On} the {Societal} {Impact} of {Open} {Foundation} {Models}},
  volume = {235},
  url = {https://proceedings.mlr.press/v235/kapoor24a.html},
  booktitle = {Proceedings of the 41st {International} {Conference} on {Machine} {Learning}},
  publisher = {PMLR},
  author = {Kapoor, Sayash and Bommasani, Rishi and Klyman, Kevin and Longpre, Shayne and Ramaswami, Ashwin and Cihon, Peter and Hopkins, Aspen K and Bankston, Kevin and Biderman, Stella and Bogen, Miranda and Chowdhury, Rumman and Engler, Alex and Henderson, Peter and Jernite, Yacine and Lazar, Seth and Maffulli, Stefano and Nelson, Alondra and Pineau, Joelle and Skowron, Aviya and Song, Dawn and Storchan, Victor and Zhang, Daniel and Ho, Daniel E. and Liang, Percy and Narayanan, Arvind},
  editor = {Salakhutdinov, Ruslan and Kolter, Zico and Heller, Katherine and Weller, Adrian and Oliver, Nuria and Scarlett, Jonathan and Berkenkamp, Felix},
  month = jul,
  year = {2024},
  pages = {23082--23104}
}

@article{lolamurtiTechAppleGoogle2026,
  title = {{Apple}, {Google} host dozens of {AI} ‘nudify’ apps like {Grok}, report finds},
  url = {https://www.cnbc.com/2026/01/27/apple-google-host-dozens-of-ai-nudify-apps-like-grok-report-finds.html},
  journal = {CNBC News},
  author = {{Lola Murti}},
  month = jan,
  year = {2026}
}

@article{jasminemithaniMinnesotaPassesNations2026,
  title = {Minnesota passes the nation’s first ban on ‘nudification’ apps},
  url = {https://19thnews.org/2026/04/minnesota-nudification-ban-ai-deepfake/},
  journal = {The 19th News},
  author = {{Jasmine Mithani}},
  month = apr,
  year = {2026}
}

@article{samanthacoleNewResearchShows2025,
  title = {New {Research} {Shows} {Deepfake} {Harassment} {Tools} {Spread} on {Social} {Media} and {Search} {Engines}},
  url = {https://www.404media.co/deepfake-tools-spread-on-social-media-research/},
  journal = {404 Media},
  author = {Cole, Samantha},
  month = oct,
  year = {2025}
}

@techreport{techtransparencyprojectAppleGoogleAre2026,
  title = {Apple and {Google} {Are} {Steering} {Users} to {Nudify} {Apps}},
  url = {https://www.techtransparencyproject.org/articles/apple-and-google-are-steering-users-to-nudify-apps},
  author = {{Tech Transparency Project}},
  month = apr,
  year = {2026}
}

@article{alexiosmantzarlisAICompanionApps,
  month = apr,
  year={2026},
  title = {{AI} companion apps are a nonconsensual nightmare},
  url = {https://indicator.media/p/ai-companion-apps-are-a-nonconsensual-nightmare},
  journal = {Indicator},
  author = {{Mantzarlis, Alexios}}
}

@misc{gracebrighamHowAICompanion2026,
  type = {Substack},
  title = {How {AI} {Companion} {Apps} {Enable} and {Encourage} {Nonconsensual} {Intimate} {Imagery}},
  url = {https://gracebrigham.substack.com/p/how-ai-companion-apps-enable-and},
  journal = {Grace Brigham},
  author = {{Grace Brigham} and {Yoshi Kohno} and {Lucy Qin}},
  month = apr,
  year = {2026}
}

@inproceedings{cretuEvaluatingConceptFiltering2026,
  title = {Evaluating {Concept} {Filtering} {Defenses} against {Child} {Sexual} {Abuse} {Material} {Generation} by {Text}-to-{Image} {Models}},
  booktitle = {Proceedings of the 2026 {IEEE} {Symposium} on {Security} and {Privacy} ({SP})},
  publisher = {IEEE},
  author = {Cretu, Ana-Maria and Kireev, Klim and Abdalla, Amro and Obinna, Wisdom and Meier, Raphael and Bargal, Sarah Adel and Redmiles, Elissa M. and Troncoso, Carmela},
  year = {2026}
}

@inproceedings{ortegaAILovesBoobies2026,
  address = {Singapore , Singapore},
  title = {{AI} {Loves} {Boobies}: {Unpacking} {Harmful} {Imaginaries} in {Community}-{Based} {Generative} {AI}},
  url = {https://dl.acm.org/doi/10.1145/3800645.3812786},
  doi = {10.1145/3800645.3812786},
  booktitle = {Proceedings of the 2026 {Designing} {Interactive} {Systems} {Conference}},
  publisher = {ACM},
  author = {Ortega, Alejandra Gómez and Van Der Maden, Willem and Lampinen, Airi and Comber, Rob and Balaam, Madeline},
  month = jun,
  year = {2026},
  pages = {158--171}
}

@inproceedings{qiweiPlatformsCrimeScene2026,
  address = {Barcelona Spain},
  title = {Platforms as {Crime} {Scene}, {Judge}, and {Jury}: {How} {Victim}-{Survivors} of {Non}-{Consensual} {Intimate} {Imagery} {Report} {Abuse} {Online}},
  url = {https://dl.acm.org/doi/10.1145/3772318.3791115},
  doi = {10.1145/3772318.3791115},
  booktitle = {Proceedings of the 2026 {CHI} {Conference} on {Human} {Factors} in {Computing} {Systems}},
  publisher = {ACM},
  author = {Qiwei, Li and Kennon, Katelyn and Bedera, Nicole and Eaton, Asia A. and Gilbert, Eric and Schoenebeck, Sarita},
  month = apr,
  year = {2026},
  pages = {1--19}
}

@techreport{thielGenerativeMLCSAM2025,
  title = {Generative {ML} and {CSAM}: {Implications} and {Mitigations}},
  url = {https://purl.stanford.edu/jv206yg3793},
  doi = {10.25740/JV206YG3793},
  institution = {Stanford Digital Repository},
  author = {Thiel, David and Stroebel, Melissa and Portnoff, Rebecca},
  year = {2023}
}

@misc{qinThoughtYouWere2026,
  title = {"{I} {Thought} {You} {Were} {The} {Uncensored} {Place}": {Norms}, {Rules}, and {Moderation} in {AI}-{Generated} {Sexual} {Content} {Communities}},
  url = {https://arxiv.org/abs/2608.13659},
  doi = {10.48550/ARXIV.2608.13659},
  publisher = {arXiv},
  author = {Qin, Lucy and Mink, Jaron and Redmiles, Elissa M.},
  month = aug,
  year = {2026},
  note = {Version Number: 1}
}

@techreport{beccabranumRapidResponseBuilding2025,
  title = {Rapid {Response}: {Building} {Victim}-{Centered} {Reporting} {Processes} for {Non}-{Consensual} {Intimate} {Imagery}},
  url = {https://cdt.org/insights/rapid-response-building-victim-centered-reporting-processes-for-non-consensual-intimate-imagery/},
  institution = {Center for Democracy \& Technology},
  author = {{Branum, Becca and Kim, Mi Yeon}},
  month = jul,
  year = {2025}
}

@book{samanthacoleHowSexChanged2022,
  title = {How {Sex} {Changed} the {Internet} and the {Internet} {Changed} {Sex}: {An} {Unexpected} {History}},
  publisher = {Workman Publishing Company},
  author = {Cole, Samantha},
  month = nov,
  year = {2022}
}

@techreport{chaynExplicitHarmsNonExplicit2026,
  title = {Explicit {Harms} of {Non}-{Explicit} {Images}},
  url = {https://reports.chayn.co/iba_pdfs/chayn-report-explicit-harms-of-non-explicit-images.pdf},
  institution = {Chayn},
  author = {{Chayn}},
  year = {2026}
}

@article{chaMyBodyNot2026,
  title = {"{My} body is not your {Porn}": {Identifying} {Trends} of {Harm} and {Oppression} through a {Sociotechnical} {Genealogy} of {Digital} {Sexual} {Violence} in {South} {Korea} {CSCW030}},
  volume = {10},
  url = {https://dl.acm.org/doi/10.1145/3788066},
  doi = {10.1145/3788066},
  number = {2},
  journal = {Proceedings of the ACM on Human-Computer Interaction},
  author = {Cha, Inha and Jang, Yeonju and Kim, Haesoo and Park, Joo Young and Park, Seora and Cheon, EunJeong},
  month = apr,
  year = {2026},
  pages = {1--32}
}

\cleardoublepage

\appendix
\cleardoublepage
\cleardoublepage
\section{\civitai{} Platform Policies}
\label{app:civitai-platform-policies}

\subsection{Self-Attestation}
\label{app:self-attestation}
\begin{figure}[h]
    \centering
    \includegraphics[width=\linewidth]{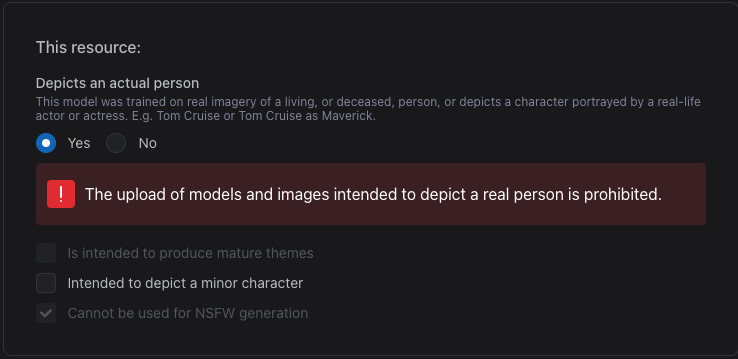}
    \caption{\small \textmd{\textbf{A required disclosure when uploading a model to \civitai. When a user confirms a model depicts likeness, they are provided the following warning and the NSFW option (but not the ``next'' button) is automatically disabled.}}}
    \label{fig:civitai-attestation}
\end{figure}

Content uploaded to \civitai{} is provided a sexual content classification, which impacts what content is visible for unauthenticated users. Additionally, per recent policy changes, \civitai{} prohibits depictions of likeness. In addition to automated scanning, they require creators to manually attest that a model is not ``intended to depict a real person'' at the upload (Figure \ref{fig:civitai-attestation}).

\subsection{Content Classification System}
\label{app:classification-system}
\begin{figure}[h]
    \centering
    \includegraphics[width=0.5\linewidth]{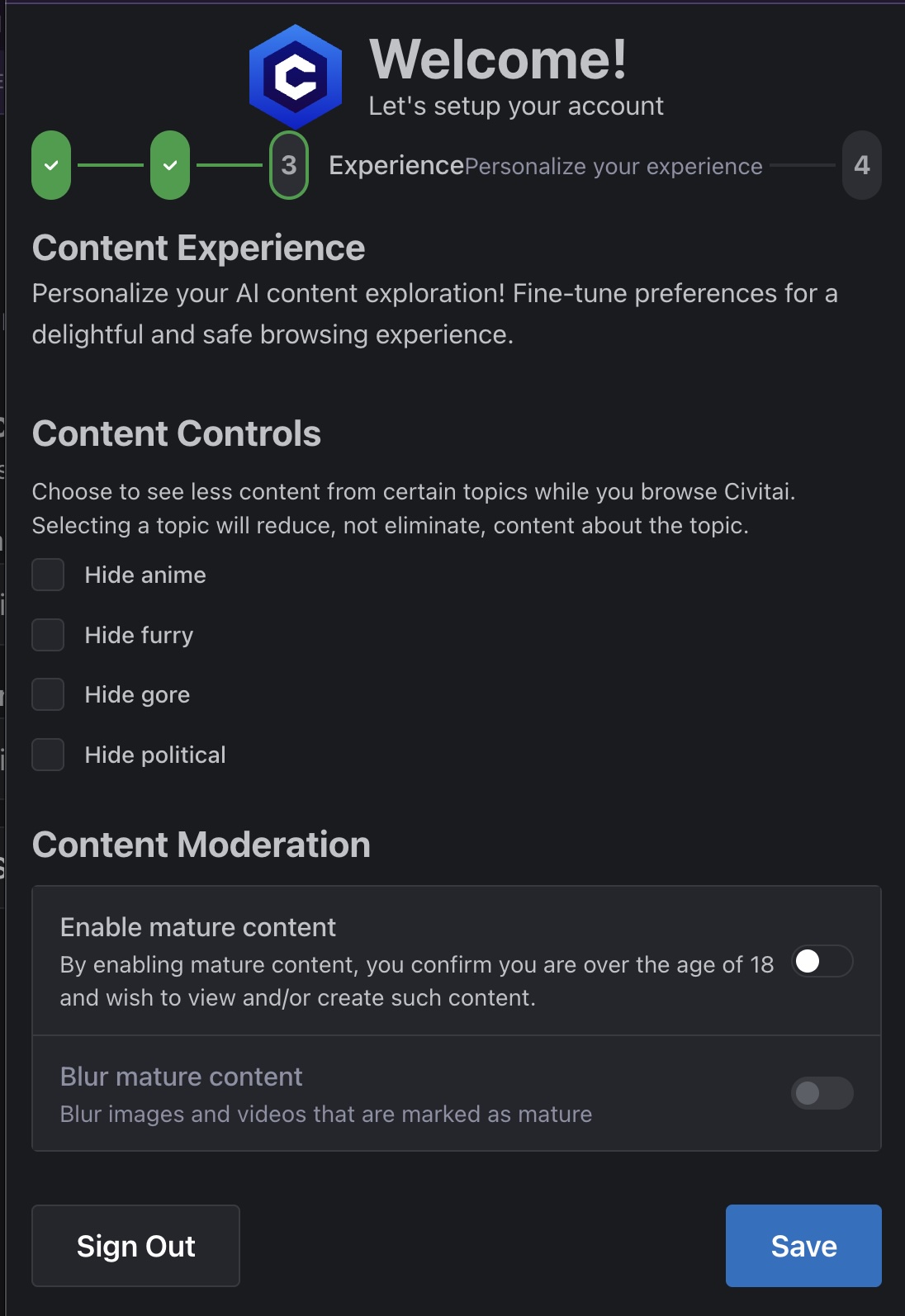}
    \caption{\small \textmd{\textbf{Onboarding Screen for \civitai{} New Users} -- ``Mature Content'' must be manually enabled.}}
    \label{civitai-content-mod}
\end{figure}

\civitai{} classifies explicit content on 5 levels: PG, PG-13, R, X, and XXX. These ratings get applied by a combination of the content creator (who can select a rating when uploading content) and the website's policy moderation, favoring the more extreme rating when there are conflicts. In the case of articles, both text and images are considered relevant for assigning a maturity level, meaning an article that does not discuss sexual content at all, but includes a nude image, would be classified as sexual. The articles collected for this paper include anything that otherwise met our filtering criteria (Section \ref{sec:methods}) and was classified as either R (defined as ``adult themes and situations'', including ''partial nudity'' or ''graphic violence''), X (defined as ``graphic nudity, genitalia, adult objects, or settings'') or XXX (defined as ``sexual acts'' and ``extremely disturbing content'')~\cite{civitaiSafety}. Users must be logged in to view R+ content. For new accounts, it is toggled off by default; Figure \ref{civitai-content-mod} depicts a screenshot where users are prompted to toggle it on. Once selected, each classification level (e.g., PG, PG-13, etc.) can be turned on or off independently.

\label{app:civitai-content-viewing}

\section{Age Risks}
\label{app:age-risks}
In addition to the introduction of sexual concepts (\sexualrisklabel) and identity concepts (\identityrisklabel), we identify two ways that components may introduce age-related risks (\agerisklabel) to technical workflows.

\apara[arisk:style-decreases-age]{Base Model: Using Models Outside Intended Style May Decrease Age}
Occasionally, guides note that fine-tuning a model may have unintended, potentially illegal consequences, particularly when used outside its intended context.
For example, guides note that Illustrious \textit{``is built specifically for anime (at the cost of being weaker for realism)''} (G264). Subsequently,
\gquote{when going `realistic' it [Illustrious] often tends to make characters appear young/er (than anticipated/expected)}~(G27).
Consistent with prior work, this may suggest that the production of AIG-CSAM is not only easy but, for intentionally offending users, even possibly accidental~\cite{goller2026systematic}.

\apara[arisk:]{Text Inputs: Prompting for ``Maturity'' Leads to Mixed Results}
As noted by (\aref{arisk:style-decreases-age}), model biases (particularly when anime-focused), may skew towards depicting younger characters. Given that \gquote{anything you leave out of the prompt is going to be generated by the model depending on its biases. In other words, `it'll do whatever it feels like'} (G92), some guides discuss prompting strategies to actively prevent CSAM, including adding terms \gquote{like ``mature female'' or ``aged up'' everywhere} (G76) or \gquote{child, underage} in the negative prompt (G169). For instance, in G27 that discusses the fictional teenage identity `Kim Possible' (from the TV show \textit{Kim Possible})
notes that they are depicted as young with no modifiers, but as older when \gquote{adult, grown\_up, mid-20s} and \gquote{minor, child, chibi, loli, lolicon, shota, shotacon} are added to the positive and negative prompts, respectively; however, they also note that precisely controlling ages, particularly across genders, is difficult: \gquote{While I can confirm [that adding mature helps]... [but] this was a change from 20 years of age to about 50 years of age} which discouraged their use~(G27).

\clearpage
\section{Codebook}
\label{appendix:codebook}
\begin{table*}[h!]
    \centering
    \small
    \begin{tabularx}{\textwidth}{X l X}
        \toprule
        \textbf{Code} & \textbf{Freq.} & \textbf{Description} \\
        \midrule
        \multicolumn{3}{l}{\textbf{Generation Workflows}} \\
        \quad Text Inputs ($\kappa=0.84$) & \textInputGuidesCount & Guides that describe workflows that start with text inputs \\
        \quad Image/Video Inputs ($\kappa=1$) & \imageInputGuidesCount & Guides that describe workflows that start with image/video inputs \\
        \quad Image Outputs ($\kappa=0.82$) & \imageOutputGuidesCount & Guides that describe workflows that generate images \\
        \quad Video Outputs ($\kappa=1$) & \videoOutputGuidesCount & Guides that describe workflows that generate videos \\
        \quad Image-based Adapters ($\kappa=1$) & \imageAdaptersGuidesCounts & Guides that describe workflows that use image-based adapters \\
        \quad Pre-Trained Adapters ($\kappa=1$) & \preTrainedAdaptersGuidesCounts & Guides that describe workflows that use pre-trained adapters \\
        \quad Pre-Trained Adapters for Sexual Content ($\kappa=1$) & \preTrainedAdaptersSexualGuidesCounts & Guides that describe workflows that use model-adapters for sexual content generation \\
        \quad Pre-Trained Adapters for Identity ($\kappa=0.83$) & \preTrainedAdaptersIdentityGuidesCounts & Guides that describe workflows that use model-adapters for generating specific identities  \\
        \midrule
        \multicolumn{3}{l}{\textbf{Training Workflows}} \\
        \quad General Training ($\kappa=0.82$) & \trainingGuidesCounts & Guides that discuss the process of modifying models \\
        \quad LoRA Training ($\kappa=1$) &  \trainingLoraGuidesCounts & Guides that discuss training LoRA, a specific popular method of pre-trained adapters known to facilitate NCII \\
        \midrule
        \multicolumn{3}{l}{\textbf{Content Type}} \\
        \quad Sexually Relevant Content ($\kappa=0.80$) & \sexuallyRelevantGuidesCount & Guides that meet by text description or visual depiction, the criteria for sexually relevant content (see \ref{sec:methods}) \\
        \quad Intimate Content ($\kappa=0.80$) & \intimateGuidesCount & Guides that meet, by text description or visual depiction, the criteria for intimate content (see \ref{sec:methods}) \\
        \quad Identifiable Personas ($\kappa=1$) & \identityGuidesCount & Guide workflows that describe how to depict or provide an example output of recognizable identity from media (e.g., Harrison Ford, Sailor Moon) \\
        \quad Photorealistic Personas ($\kappa$=$0.87$) & \photorealisticGuidesCount & guides that describe how to depict or provide an example of outputs that are indistinguishable from real photographs \\
    \bottomrule
    \end{tabularx}
    \caption{\textbf{Primary Codebook} with code frequencies and related descriptions}
    \label{tab:placeholder}
\end{table*}

\definecolor{closedrose}{RGB}{253,237,237}
\definecolor{openblue}{RGB}{235,244,252}
\definecolor{softline}{gray}{0.75}

\begin{table*}[htbp]
\centering
\footnotesize
\begin{tabular}{l p{2.6cm} p{2.6cm} p{3.2cm} r l}
\toprule
\textbf{Open/Closed} & \textbf{Foundation Model} & \textbf{Fine-Tuning} & \textbf{Derivative} & \textbf{Count} & \textbf{ Output Types} \\
\midrule
\rowcolor{openblue}
\multirow[t]{38}{*}{\textbf{Open}} & \multirow[t]{20}{*}{\textbf{Stable Diffusion}} &  &  & \textbf{158} &  Image \\
\rowcolor{openblue} &  & Waifu Diffusion 14 &  & 6 & Text \\
\rowcolor{openblue} &  & CyberRealistic &  & 2 & Image  \\
\rowcolor{openblue} &  & bigASP &  & 2 & Image \\
\rowcolor{openblue} &  & DTFMix* &  & 1 & Image \\
\rowcolor{openblue} &  & StellaratorMix* &  & 1 & Image \\
\rowcolor{openblue} &  & VXP XL (Hyper) &  & 1 & Image \\
\rowcolor{openblue} &  & \multirow[t]{8}{*}{\textbf{Illustrious}} &  & \textbf{69} &  Image \\
\rowcolor{openblue} &  &  & NoobAI & 9 & Image \\
\rowcolor{openblue} &  &  & WAI-NSFW-Illustrious-SDXL & 8 & Image \\
\rowcolor{openblue} &  &  & Hassaku & 3 & Image \\
\rowcolor{openblue} &  &  & Illustrij & 3 & Image \\
\rowcolor{openblue} &  &  & Model-Merges & 1 & Image \\
\rowcolor{openblue} &  &  & Additional Fine-Tunings & 21 & Image \\
\rowcolor{openblue} &  & \multirow[t]{5}{*}{\textbf{Pony}} &  & \textbf{44} & Image \\
\rowcolor{openblue} &  &  & CyberRealistic Pony & 5 & Image \\
\rowcolor{openblue} &  &  & FissionMix* & 1 & Image \\
\rowcolor{openblue} &  &  & Model-Merges & 1 & Image \\
\rowcolor{openblue} &  &  & Additional Fine-Tunings & 21 & Image \\
\arrayrulecolor{softline}\cmidrule(l){2-6}
\rowcolor{openblue} & \multirow[t]{3}{*}{\textbf{Flux}} &  &  & 12 & Image \\
\rowcolor{openblue} &  &  & Instagram Women & 1 & Image \\
\rowcolor{openblue} &  &  & Big Love Photo1 & 1 &  Image\\
\rowcolor{openblue} & Qwen &  &  & 4 &  Image \\
\rowcolor{openblue} & Z-Image Turbo &  &  & 1 & Image \\
\rowcolor{openblue} & PoissonMix* &  &  & 1 & Image \\
\rowcolor{openblue} & TernaryMix* &  &  & 1 &  Image \\
\rowcolor{openblue} & BremsstrahlungMix* &  &  & 1 & Image \\
\arrayrulecolor{softline}\cmidrule(l){2-6}
\rowcolor{openblue} & Wan &  &  & 18 & Video \\
\rowcolor{openblue} & Hunyuan Video &  &  & 9 &  Video \\
\rowcolor{openblue} & Anisora &  &  & 1 & Video \\
\rowcolor{openblue} & LTX Video (LTXV) &  &  & 2 & Video \\
\arrayrulecolor{softline}\cmidrule(l){2-6}
\rowcolor{openblue} & Florence2 &  &  & 3 & Text \\
\rowcolor{openblue} & MoonDream &  &  & 1 &  Text \\
\rowcolor{openblue} & Joy Caption &  &  & 2 & Text \\
\rowcolor{openblue} & Gemma2 &  &  & 1 & Text \\
\rowcolor{openblue} & Dolphin &  &  & 1 & Text \\
\rowcolor{openblue} & DeepSeek &  &  & 1 & Text \\
\midrule
\rowcolor{closedrose}
\multirow[t]{11}{*}{Closed} & ChatGPT &  &  & 10 & Text \\
\rowcolor{closedrose} & Grok-Text &  &  & 3 & Text \\
\rowcolor{closedrose} & Gemini &  &  & 2 & Text \\
\arrayrulecolor{softline}\cmidrule(l){2-6}
\rowcolor{closedrose} & Midjourney &  &  & 4 & Image \\
\rowcolor{closedrose} & NanoBanana &  &  & 3 & Image \\
\rowcolor{closedrose} & Seedream &  &  & 1 &  Image\\
\rowcolor{closedrose} & Dall-E &  &  & 1 &  Image\\
\arrayrulecolor{softline}\cmidrule(l){2-6}
\rowcolor{closedrose} & Kling &  &  & 4 & Video \\
\rowcolor{closedrose} & PixVerse &  &  & 1 & Video \\
\rowcolor{closedrose} & Mochi &  &  & 1 & Video \\
\arrayrulecolor{softline}\cmidrule(l){2-6}
\rowcolor{closedrose} & Grok-Image/Video &  &  & 2 & Image/Video \\
\bottomrule
\end{tabular}
\caption{\textbf{All Models} observed across the guides. \textbf{*} indicates a model has since been removed from \civitai.}
\label{tab:all-models}
\end{table*}

\end{document}